# Conformal Transformation Electromagnetics based on Schwarz-Christoffel Mapping for the Synthesis of Doubly-Connected Metalenses

M. Salucci, *Member, IEEE*, F. Boulos, A. Polo, *Member, IEEE*, and G. Oliveri, *Senior Member, IEEE*

***Abstract*—An innovative transformation electromagnetics (*TE*) paradigm, which leverages on the Schwarz-Christoffel (*SC*) theorem, is proposed to design effective and realistic field manipulation devices (*FMD*). Thanks to the conformal property, such a *TE* design method allows one to considerably mitigate the anisotropy of the synthesized metalenses (i.e., devices with artificially-engineered materials covering an antenna to modify its radiation features) with respect to those yielded by competitive state-of-the-art *TE* techniques. Moreover, devices with doubly-connected contours, thus including masts with arbitrary sections and lenses with holes/forbidden regions in which the material properties cannot be controlled, can be handled. A set of numerical experiments is presented to assess the features of the proposed method in terms of field-manipulation capabilities and complexity of the lens material also in a comparative fashion.**

## I. Introduction and Rationale

In the last few years, transformation electromagnetics (*TE*) has emerged as a powerful paradigm for the design of innovative electromagnetic (*EM*) devices with unprecedented field manipulating capabilities [1]-[3]. The key idea of *TE* is that the wave propagation properties of a reference radiating device can be mimicked by another one with a different shape, but coated with a properly designed metalens (i.e., a device whose constituent materials are artificially-engineered/not already available in nature to achieve user-defined/unconventional radiation features [1]) [1]. Towards this end, suitable coordinate transformations are applied in the reference space in which the antenna operates to (*i*) derive new material specifications starting from transformed geometrical parameters and to (*ii*) control the propagation of *EM* waves within the surrounding environment [1]. The generality and the flexibility of such an idea have been demonstrated in several applications and *TE*-based devices have been implemented across a broad range of frequencies and scenarios, including carpet cloaks [4], optic graded-index metalenses [5][6], conformal arrays and radiators [7]-[10], multibeam antennas [11], compressed Luneburg lenses [12][13], flat reflectors [14][15], waveguide couplers [16], focusing lenses [17][18], miniaturized antennas and arrays [19][20], and compact horns [21]. For instance, it has been already demonstrated that *TE* can enable the design of *L*-shaped arrays coated with metamaterial lenses able to generate, outside the lens support, the same field distribution of circular layouts radiating in the free space, thus avoiding scan loss/beam distortion issues [7]. The popularity of *TE* and its extensions [22][23] are motivated by several factors such as the strong theoretical foundations, which rely on the invariance of Maxwell's equations under coordinate transformations [2], and the availability of *closed-form* exact expressions for the synthesis of the metalenses once a suitable coordinate mapping between the reference geometry and the designed one is available [1]. On the other hand, it is worth pointing out that the derivation/computation of the mapping functions is often a challenging task from both the methodological and the practical viewpoint since (*i*) handling non-canonical shapes has no trivial and general-purpose solutions and (*ii*) yielding isotropic (or low anisotropy) lenses is generally difficult even for elementary geometries [1]. In this framework, *Quasi-Conformal TE* (*QCTE*) methods deal with non-canonical shapes yielding a reduced material anisotropy [1][7], but their direct (i.e., without intermediate steps and approximations) application is limited to simply-connected domains [7]. Therefore, the *QCTE* design of field manipulation devices including holes/empty regions, which are of potential interest for several applications such as mast-mounted *5G* base stations and/or radar antennas, is actually inefficient.

To overcome this issue, an innovative class of *TE* design strategies is here proposed. Thanks to the Schwarz-Christoffel (*SC*) theorem [24], a conformal mapping technique is customized to the solution of the *TE* problem to derive a new synthesis approach that (*i*) ideally avoids/minimizes the anisotropy of the metalenses and (*ii*) successfully handles arbitrary doubly-connected domains (i.e., lenses/devices including holes and/or forbidden regions). Such an approach is mainly motivated by the following reasons: (*i*) the exploitation of *SC* mapping is

Manuscript received May 27, 2019
This work benefited from the networking activities carried out within the Project Cloaking Metasurfaces for a New Generation of Intelligent Antenna systems (MANTLES)" funded by the Italian Ministry of Education, University, and Research within the PRIN2017 Program, within the Project SNATCH funded by the Italian Ministry of Foreign Affairs and International Cooperation, Directorate General for Cultural and Economic Promotion and Innovation (2017-2019), and within the Project "SMARTOUR - Piattaforma Intelligente per il Turismo" (Grant no. SCN_00166) funded by the Italian Ministry of Education, University, and Research within the Program "Smart cities and communities and Social Innovation".
M. Salucci, F. Boulos, A. Polo, and G. Oliveri are with the ELEDIA Research Center (ELEDIA@UniTN - University of Trento), Via Sommarive 9, 38123 Trento - Italy (e-mail: {marco.salucci, federico.boulos, alessandro.polo.1, giacomo.oliveri}@unitn.it)
M. Salucci and G. Oliveri are also with the ELEDIA Research Center (ELEDIA@L2S - UMR 8506), 3 rue Joliot Curie, 91192 Gif-sur-Yvette - France (e-mail: {marco.salucci, giacomo.oliveri}@l2s.centralesupelec.fr)
[1]It is worth remarking that the word "metalens" is here adopted to indicate *TE*-synthesized devices and it is not limited to the concept of metasurfaces.

expected to considerably improve the lens performance and the design flexibility with respect to standard *QCTE* approaches because of the inherently conformal nature of the resulting transformation [24], (*ii*) effective numerical implementations of the *SC* mapping formula [25] can be customized to the problem of interest also when very complex lens profiles are at hand, (*iii*) the resulting theory is a generalization of previously-derived conformal transformation approaches since all conformal transformations with known analytic forms are also *SC* maps, sometimes disguised by a change of variables [24].

As for the key methodological and innovative contributions of this paper, they include (*i*) for the first time to the best of the authors' knowledge, the customization and the application of *SC* mapping strategies to the solution of *TE* design problems, (*ii*) the introduction of a *TE* technique suitable for doubly-connected domains, thus the possibility of designing lenses with holes or forbidden regions, and (*iii*) the derivation of a set of operative guidelines for the effective exploitation of the proposed synthesis approach.

The outline of the paper is as follows. The formulation of the design problem at hand, within the *TE* framework, is presented in Sect. II. Section III details the customization of the *SC* mapping to the solution of the arising conformal *TE* formulation. A set of representative numerical results, from a broad numerical validation, is then illustrated to assess the effectiveness and the potentialities of the proposed *SC-CTE* design technique also in comparison with state-of-the-art *QCTE* methodologies (Sect. IV). Finally, some conclusions and remarks are drawn (Sect. V).

## II. Formulation of the Design Problem within the *TE* Framework

Let us consider the problem of synthesizing the dielectric properties [i.e., the relative permittivity and the permeability tensors $\overline{\overline{\varepsilon}}(\mathbf{r})$ and $\overline{\overline{\mu}}(\mathbf{r})$] of a doubly-connected metalens with user-defined shape $\Gamma$ coating an $N$-element array [Fig. 1(*a*)] of time-harmonic [2] current line sources $\mathbf{J}_n(\mathbf{r}) = J_n \delta(\mathbf{r} - \mathbf{r}_n)\widehat{\mathbf{z}}$, $n = 1, ..., N$, so that the resulting device [i.e., the ensemble of the $N$-element array and the metalens, which will be indicated - in the following - as *field manipulation device* (*FMD*)] radiates an electric, $\mathbf{e}$, and a magnetic, $\mathbf{h}$, field distributions that comply with user-specified *target* fields, $\mathbf{e}^*$ and $\mathbf{h}^*$, outside the *FMD* volume ($\mathbf{r} \notin \Gamma$)

$$\begin{cases} \mathbf{e}(\mathbf{r}) = \mathbf{e}^*(\mathbf{r}) \\ \mathbf{h}(\mathbf{r}) = \mathbf{h}^*(\mathbf{r}) \end{cases} \quad \mathbf{r} \notin \Gamma. \tag{1}$$

Then, let us consider an auxiliary space described by the coordinate system $\mathbf{r}^* \triangleq (x^*, y^*, z^*)$ and let us assume that the target fields $\mathbf{e}^*$ and $\mathbf{h}^*$ are the result of the *EM* interactions between $N$ electric current line sources, $\{\mathbf{J}_n^*(\mathbf{r}^*) = J_n^* \delta(\mathbf{r}^* - \mathbf{r}_n^*)\widehat{\mathbf{z}}^*;\ n = 1, ..., N\}$ ($J_n^* = e^{-j\frac{2\pi}{\lambda}\left[\nu_i^* \cos\left(\frac{2\pi}{N}[n-1]\right)\cos(\varphi_0) + \nu_i^* \sin\left(\frac{2\pi}{N}[n-1]\right)\sin(\varphi_0)\right]}$, $\varphi_0$ being the beam-steering direction), and a doubly-connected *annular* region $\Gamma^*$, with inner/outer radius equal to $\nu_i^*/\nu_o^*$ [Fig. 1(*b*)], characterized by the relative permittivity and the permeability tensors $\overline{\overline{\varepsilon}}^*(\mathbf{r}^*)$ and $\overline{\overline{\mu}}^*(\mathbf{r}^*)$ ($\mathbf{r}^* \in \Gamma^*$),[3] respectively. It means that

$$\begin{cases} \mathbf{e}^*(\mathbf{r}) \triangleq \mathbf{e}^*(\mathbf{r}^*)\rfloor_{\mathbf{r}^*=\mathbf{r}} \\ \mathbf{h}^*(\mathbf{r}) \triangleq \mathbf{h}^*(\mathbf{r}^*)\rfloor_{\mathbf{r}^*=\mathbf{r}} . \end{cases} \tag{2}$$

[2] The time dependence $\exp(j2\pi f t)$ is assumed and omitted hereinafter.

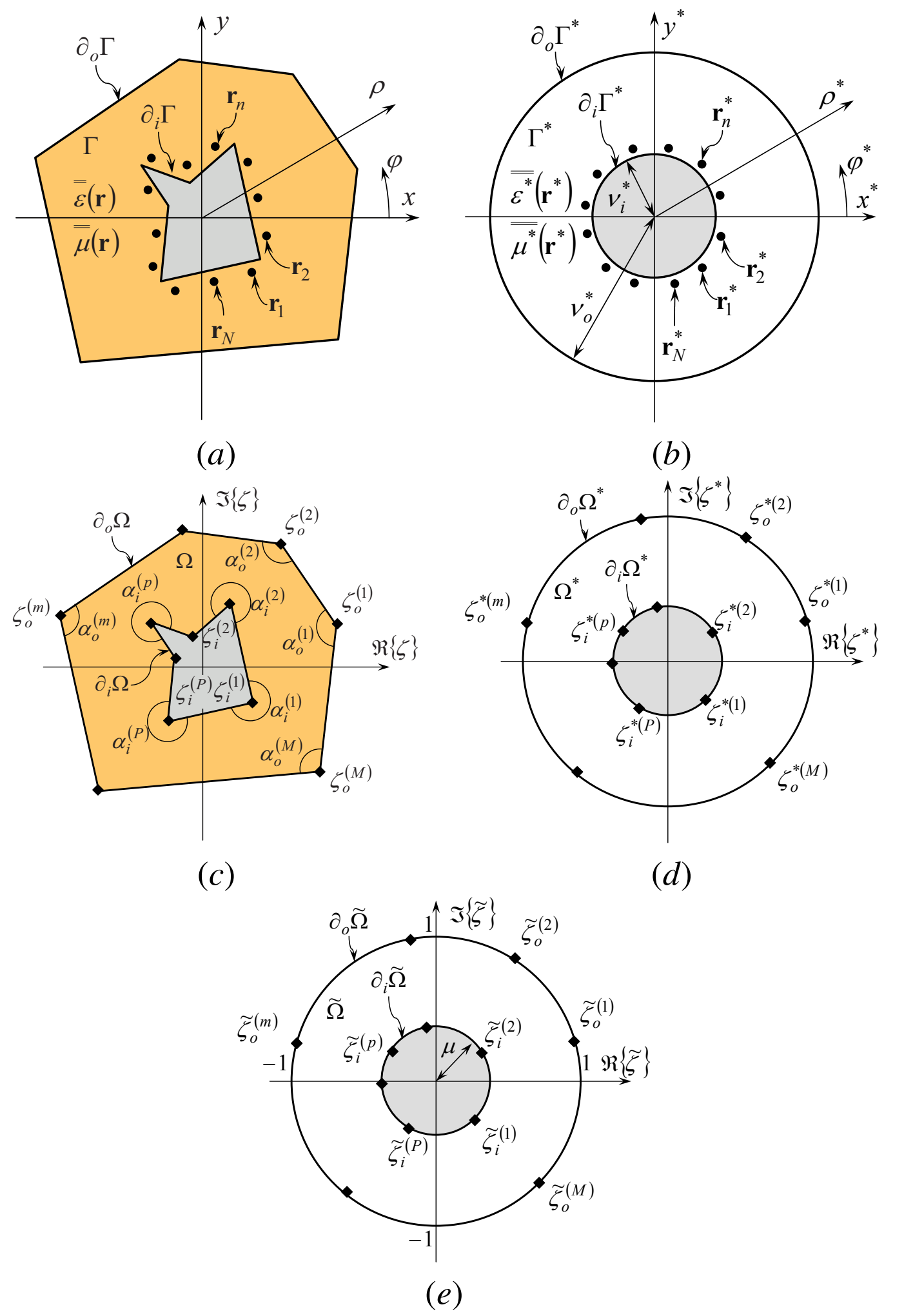

Figure 1. *Problem Geometry*. Sketches of (*a*) the doubly-connected metalens with user-defined shape $\Gamma$, (*b*) the doubly-connected reference annular region $\Gamma^*$, (*c*) the two auxiliary complex spaces (*d*) $\Omega$ ($\zeta$: $\Gamma \to \Omega$) and (*e*) $\Omega^*$ ($\zeta^*$: $\Gamma^* \to \Omega^*$), and (*e*) the anulus region $\widetilde{\Omega}$ ($\chi$: $\Omega^* \to \widetilde{\Omega}$; $\psi$: $\widetilde{\Omega} \to \Omega$) with internal and external radius equal to $\mu$ [$\mu \in (0,1)$] and to 1, respectively.

By adopting the *TE* guidelines [1], the solution of this problem can be found if a suitable mapping between the two coordinate systems, $\mathbf{r}^*$ and $\mathbf{r}$, is derived by defining the *transformation function* $\boldsymbol{\tau}$

$$\mathbf{r} = \boldsymbol{\tau}(\mathbf{r}^*) \triangleq [\tau_x(\mathbf{r}^*), \tau_y(\mathbf{r}^*), \tau_z(\mathbf{r}^*)] \tag{3}$$

that fits the contour mapping equation

$$\boldsymbol{\tau}(\mathbf{r}^*)\rfloor_{\mathbf{r}^* \in \{\partial_i\Gamma^* \cup \partial_o\Gamma^*\}} = \{\partial_i\Gamma \cup \partial_o\Gamma\} \tag{4}$$

$\partial_i$ and $\partial_o$ being the *inner* and the *outer* contour operators of a generic doubly-connected region. Indeed, because of the form-invariant nature of the Maxwell's equations under spatial transformations and by exploiting the "material interpretation" concept [1], one can prove that (1) is exactly satisfied by a *FMD* defined by the following conditions:

$$\begin{cases} \mathbf{r}_n = \boldsymbol{\tau}\left(\mathbf{r}_n^*\right) \\ J_n = J_n^* \end{cases} \qquad n = 1, ..., N \tag{5}$$

[3] Without loss of generality, $\Gamma^*$ can be always defined as an annular region as $\Gamma^* \triangleq \{\nu_i \leq \mathbf{r}^* \leq \nu_o\}$ such that $\Gamma^* \supset \{\mathbf{r}_n^*, n = 1, ..., N\}$, $\Gamma^* \supset \text{supp}\left[\overline{\overline{\varepsilon}}^*(\mathbf{r}^*) - 1\right]$, and $\Gamma^* \supset \text{supp}\left[\overline{\overline{\mu}}^*(\mathbf{r}^*) - 1\right]$, supp $[\cdot]$ being the *support* operator.

(*Source Mapping* Condition)

$$\begin{cases} \overline{\overline{\varepsilon}}\left(\mathbf{r}\right) = \left.\frac{\mathcal{J}[\boldsymbol{\tau}(\mathbf{r}^*)]\overline{\overline{\varepsilon}}(\mathbf{r}^*)\mathcal{J}^T[\boldsymbol{\tau}(\mathbf{r}^*)]}{\det\{\mathcal{J}[\boldsymbol{\tau}(\mathbf{r}^*)]\}}\right\rfloor_{\mathbf{r}^*=\mathbf{r}} \\ \overline{\overline{\mu}}\left(\mathbf{r}\right) = \left.\frac{\mathcal{J}[\boldsymbol{\tau}(\mathbf{r}^*)]\overline{\overline{\mu}}(\mathbf{r}^*)\mathcal{J}^T[\boldsymbol{\tau}(\mathbf{r}^*)]}{\det\{\mathcal{J}[\boldsymbol{\tau}(\mathbf{r}^*)]\}}\right\rfloor_{\mathbf{r}^*=\mathbf{r}} \end{cases} \quad \text{if } \mathbf{r} \in \Gamma \tag{6}$$

(*Material Mapping* Condition)

where $\det\{\cdot\}$ stands for the determinant, $\mathcal{J}[\cdot]$ is the Jacobian tensor, and $\cdot^T$ is the transpose operator.

Accordingly, the original *FMD* synthesis problem can be formulated as the following *transformation function* design:

> *TE-Based Design Problem* (*TEDP*) - Given $\partial_i\Gamma$, $\partial_o\Gamma$, $\partial_i\Gamma^*$, and $\partial_o\Gamma^*$, find the transformation function $\boldsymbol{\tau}$ ($\boldsymbol{\tau}$: $\Gamma^* \to \Gamma$) such that the contour mapping equation (4) holds true.

Indeed, once $\boldsymbol{\tau}$ is known, the lens dielectric properties and the resulting array configuration that radiate exactly the target field (1) can be simply found by using (5)-(6) given the reference line sources configuration, $\{\mathbf{J}_n^*(\mathbf{r}^*);\ n = 1, ..., N\}$, and the reference material tensors $\overline{\overline{\varepsilon}}^*(\mathbf{r}^*)$ and $\overline{\overline{\mu}}^*(\mathbf{r}^*)$, $\mathbf{r}^* \in \Gamma^*$.

In principle, any mapping function solving the above *TE* problem can be employed to deduce a suitable lens and array design. However, the transformation function $\boldsymbol{\tau}$ often turns out to be non-smooth even for simple geometrical configurations, unless additional constraints are forced. Generally, the result is that the synthesized lens has a strong anisotropy or unpractical $\overline{\overline{\varepsilon}}(\mathbf{r})$ and $\overline{\overline{\mu}}(\mathbf{r})$ distributions [26]. Moreover, analytically expressing $\partial_i\Gamma$, $\partial_o\Gamma$, $\partial_i\Gamma^*$, and $\partial_o\Gamma^*$ is usually a hard task with no obvious solutions except for canonical shapes [26]. Therefore, a more effective reformulation of the *transformation function* design problem is proposed.

## III. *CTE*-Synthesis by Schwarz-Christoffel Mapping

To address the *TE-Based Design Problem*, while yielding a minimum-complexity lens [26], the anisotropy of the dielectric parameters in (6) must be minimized. As a matter of fact, non-negligible deviations from the isotropic behavior, which usually arise in *TE* problems, are very difficult in practice to be accurately matched [26] with simple manufacturing processes. Otherwise, microstructure resonances are required, but the resulting bandwidth is severely limited [26]. To overcome these issues, the synthesis problem is recast to the following *conformal transformation electromagnetics* (*CTE*) *mapping* one:

> *CTE-Based Design Problem* (*CTEP*) - Given $\partial_i\Gamma$, $\partial_o\Gamma$, $\partial_i\Gamma^*$, and $\partial_o\Gamma^*$, find the transformation function $\boldsymbol{\tau}$ that identifies a *conformal* mapping between the coordinate systems, $\Gamma^* \to \Gamma$, and it also fits the contour mapping equation (4).

As a matter of fact, the anisotropy of the metalenses from a *CTE* transformation is expected to be minimized [26] since a conformal mapping preserves the angles as well as the orientation between any directed curves through each $\mathbf{r}^*$ in $\Gamma^*$ [24]. On the other hand, one should consider that the Liouville's theorem [27] states that 3D conformal mappings exist only for few geometries that do not comply with generic definitions of $\partial_i\Gamma$, $\partial_o\Gamma$, $\partial_i\Gamma^*$, and $\partial_o\Gamma^*$. However, a *2D* version of the *CTEP* can be defined since the mapping of the $z$ variable for the bi-dimensional nature of the scenario at hand [Fig. 1(*a*)], no deformation taking place along the $\widehat{\mathbf{z}}$ direction, is known to be

$$\tau_z\left(\mathbf{r}^*\right) = z^* \tag{7}$$

and the actual unknown dependences reduce to $\tau_x\left(\mathbf{r}^*\right) \triangleq \tau_x\left(x^*, y^*\right)$ and $\tau_y\left(\mathbf{r}^*\right) \triangleq \tau_y\left(x^*, y^*\right)$ only, while the Jacobian tensor in (6) is given by [28]

$$\mathcal{J}\left[\boldsymbol{\tau}\left(\mathbf{r}^*\right)\right] = \begin{bmatrix} \frac{\partial\tau_x(x^*,y^*)}{\partial x^*} & \frac{\partial\tau_x(x^*,y^*)}{\partial y^*} & 0 \\ \frac{\partial\tau_y(x^*,y^*)}{\partial x^*} & \frac{\partial\tau_y(x^*,y^*)}{\partial y^*} & 0 \\ 0 & 0 & 1 \end{bmatrix}. \tag{8}$$

According to the conformal mapping theory [24], the arising *2D* problem can be more effectively solved in the auxiliary *complex* spaces $\Omega$ [Fig. 1(*c*)] and $\Omega^*$ [Fig. 1(*d*)] defined by the following coordinate transformations ($\zeta$: $\Gamma \to \Omega$; $\zeta^*$: $\Gamma^* \to \Omega^*$)

$$\begin{cases} \zeta \triangleq x + jy \\ \zeta^* \triangleq x^* + jy^* \end{cases} \tag{9}$$

through an auxiliary transformation function $\xi$

$$\xi\left(\zeta^*\right) \triangleq \tau_x\left(\mathbf{r}^*\right) + j\tau_y\left(\mathbf{r}^*\right) \qquad \mathbf{r}^* = \left(\mathbb{R}\left\{\zeta^*\right\}, \mathbb{I}\left\{\zeta^*\right\}\right) \tag{10}$$

so that the original *CTEP* turns out to be equivalent to its complex counterpart

> *Complex CTE-Based Design Problem* (*CCTEP*) - Given $\partial_o\Omega$, $\partial_o\Omega$, $\partial_i\Omega^*$, and $\partial_o\Omega^*$, find the auxiliary transformation function $\xi$ such that the corresponding complex contour mapping equation
>
> $$\xi\left(\zeta^*\right)\rfloor_{\zeta^* \in \{\partial_i\Omega^* \cup \partial_o\Omega^*\}} = \{\partial_i\Omega \cup \partial_o\Omega\} \tag{11}$$
>
> is satisfied, $\xi$ ($\xi$: $\Omega^* \to \Omega$) being a complex *conformal* mapping function.

In (11), $\partial_j\Omega$ ($\partial_j\Omega \triangleq \{\zeta \in \Omega : (\mathbb{R}\{\zeta\}, \mathbb{I}\{\zeta\}) \in \partial_j\Gamma\}$) and $\partial_j\Omega^*$ ($\partial_j\Omega^* \triangleq \{\zeta^* \in \Omega^* : (\mathbb{R}\{\zeta^*\}, \mathbb{I}\{\zeta^*\}) \in \partial_j\Gamma^*\}$) $j \in \{i, o\}$ are the contours of the regions $\Omega$ and $\Omega^*$, which are the counterparts, in the complex plane, of the doubly-connected regions $\Gamma$ and $\Gamma^*$, respectively [Fig. 1(*c*) vs. Fig. 1(*a*) and Fig. 1(*d*) vs. Fig. 1(*b*)].

It is worthwhile noticing that, through the *CCTEP* formulation, the solution process has been shifted from the search for a two-dimensional real-valued transformation function, $\boldsymbol{\tau}\left(\mathbf{r}^*\right) \triangleq \left[\tau_x\left(x^*, y^*\right), \tau_y\left(x^*, y^*\right), z^*\right]$ (*CTEP*), to that of a *1D* complex auxiliary one, $\xi\left(\zeta^*\right)$. As a matter of fact, the solution to the original *CTEP* problem can be found by inverting (10) as follows

$$\begin{cases} \tau_x\left(\mathbf{r}^*\right) = \mathbb{R}\left\{\xi\left(\zeta^*\right)\right\}\rfloor_{\zeta^*=x^*+jy^*} \\ \tau_y\left(\mathbf{r}^*\right) = \mathbb{I}\left\{\xi\left(\zeta^*\right)\right\}\rfloor_{\zeta^*=x^*+jy^*} \end{cases} \quad \mathbf{r}^* \in \Gamma^*. \tag{12}$$

Thanks to this complex-plane reformulation, fundamental results from complex analysis can be exploited to find the mapping between $\Omega$ and $\Omega^*$. More specifically, since the Riemann Mapping (*RM*) theorem [24] guarantees that any doubly-connected domain $\Omega$ is conformally equivalent to an *annulus* $\widetilde{\Omega}$ in the complex plane ($\mathbb{R}\left\{\widetilde{\zeta}\right\}$,$\mathbb{I}\left\{\widetilde{\zeta}\right\}$) with internal radius $\mu$ [$\mu \in (0,1)$] and unitary external radius [Fig. 1(*e*)], the *CCTEP* turns out to be univocally described as the composition of two conformal mappings

$$\xi\left(\zeta^*\right) = \psi\left(\widetilde{\zeta}\right)\Big|_{\widetilde{\zeta}=\chi(\zeta^*)} \tag{13}$$

defined by the conformal mapping functions $\psi\left(\widetilde{\zeta}\right)$ and $\chi\left(\zeta^*\right)$ between the domains $\widetilde{\Omega} \to \Omega$ and the domains $\Omega^* \to \widetilde{\Omega}$, respectively. While the former mapping $\psi$ certainly exists, as a consequence of the *RM* theorem, the Schottky's theorem [29] states that the latter one, $\chi$, can be defined if and only if

$$\frac{\nu_o^*}{\nu_i^*} = \frac{1}{\mu} \tag{14}$$

$\frac{1}{\mu}$ being the *conformal modulus* of $\Omega$ [24]. Let us now observe that the choice of $\nu_o^*$ is arbitrary in our synthesis problem, since it represents the outer boundary of the reference doubly-connected *annular* region $\Gamma^*$ that can be arbitrarily extended [Fig. 1(*b*)]. Therefore, (14) holds true when setting the value $\nu_o^* = \frac{\nu_i^*}{\mu}$ and deducing that [29]

$$\chi\left(\zeta^*\right) = \frac{\mu}{\nu_i^*}\zeta^*. \tag{15}$$

As for the computation of $\psi\left(\widetilde{\zeta}\right)$, while the *RM* theorem does not provide rules for the definition of the mapping between $\Omega$ and $\widetilde{\Omega}$ [24], otherwise the *SC* theorem gives explicit formulas useful towards this end [24][25][30]. More specifically, the following numerical procedure can be inferred. Let the inner and the outer contours of $\Omega$ be discretized in $P$ internal vertexes, $\{\zeta_i^{(p)} \in \partial_i\Omega;\ p = 1, ..., P\}$, and $M$ external vertexes, $\{\zeta_o^{(m)} \in \partial_o\Omega;\ m = 1, ..., M\}$, respectively [Fig. 1(*c*)]. The conformal mapping function $\psi$ is then computed by applying the explicit *SC* formula [24][25][30]

$$\psi\left(\widetilde{\zeta}\right) = \zeta_i^{(1)} + C\int_{\widetilde{\zeta}_i^{(1)}}^{\widetilde{\zeta}} \mathcal{Q}(s)\,\mathrm{d}s \tag{16}$$

where the complex-plane integral is path-independent and the *SC* integrand $\mathcal{Q}(s)$ is given by

$$\mathcal{Q}(s) \triangleq \prod_{m=1}^{M}\left[\Theta\left(\frac{s}{\mu\widetilde{\zeta}_o^{(m)}}\right)\right]^{\beta_o^{(m)}} \prod_{p=1}^{P}\left[\Theta\left(\frac{\mu s}{\widetilde{\zeta}_i^{(p)}}\right)\right]^{\beta_i^{(p)}}, \tag{17}$$

the $\Theta$-function being defined as [25]

$$\Theta(s) \triangleq \prod_{d=1,3,5}^{\infty}\left[\left(1-\mu^d s\right)\left(1-\frac{\mu^d}{s}\right)\right], \tag{18}$$

and the *turning* parameters $\beta_o^{(m)}$, $m = 1, ..., M$, and $\beta_i^{(p)}$, $p = 1, ..., P$, being equal to [30]

$$\begin{cases} \beta_o^{(m)} \triangleq \frac{\alpha_o^{(m)}}{\pi} - 1 & m = 1, ..., M \\ \beta_i^{(p)} \triangleq \frac{\alpha_i^{(p)}}{\pi} - 1 & p = 1, ..., P, \end{cases} \tag{19}$$

where $\alpha_o^{(m)}$ ($\alpha_i^{(p)}$) is the $m$-th ($p$-th) vertex angle of the polygonal discretization of $\partial_o\Omega$ ($\partial_i\Omega$) at the corners $\zeta_o^{(m)}$ ($\zeta_i^{(p)}$) measured counterclockwise from the interior of $\Omega$ [$m = 1, ..., M$ ($p = 1, ..., P$) - Fig. 1(*c*)], respectively. Moreover, the (unknown) outer, $\{\widetilde{\zeta}_o^{(m)} \triangleq \psi^{-1}\left(\zeta_o^{(m)}\right);\ m = 1, ..., M\}$, and the inner, $\{\widetilde{\zeta}_i^{(p)} \triangleq \psi^{-1}\left(\zeta_i^{(p)}\right);\ p = 1, ..., N\}$, *prevertexes*, the internal radius $\mu$ [Fig. 1(*e*)], and the complex constant $C$ are the *SC* "accessory parameters" to be numerically determined as shown in the Appendix.

Once determined from the minimization of (29), these *SC* parameters are substituted in (16), which is then combined with (15) to yield (13). As a result, the *Complex CTE* problem turns out to be solved by the following mapping function

$$\xi\left(\zeta^*\right) = \zeta_i^{(1)} + C\int_{\psi^{-1}\left(\zeta_i^{(1)}\right)}^{\frac{\mu}{\nu_i}\zeta^*} \mathcal{Q}(s)\,\mathrm{d}s, \tag{20}$$

which is used in (12) to obtain the final *2D SC-CTE* solution

$$\begin{cases} \tau_x\left(\mathbf{r}^*\right) = x_i^{(1)} + \mathbb{R}\left\{C\int_{\psi^{-1}\left(x_i^{(1)}+jy_i^{(1)}\right)}^{\frac{\mu}{\nu_i}(x^*+jy^*)} \mathcal{Q}(s)\,\mathrm{d}s\right\} \\ \tau_y\left(\mathbf{r}^*\right) = y_i^{(1)} + \mathbb{I}\left\{C\int_{\psi^{-1}\left(x_i^{(1)}+jy_i^{(1)}\right)}^{\frac{\mu}{\nu_i}(x^*+jy^*)} \mathcal{Q}(s)\,\mathrm{d}s\right\} \end{cases} \mathbf{r}^* \in \Gamma^* \tag{21}$$

where $x_i^{(p)} = \mathbb{R}\left\{\zeta_i^{(p)}\right\}$ ($x_o^{(m)} = \mathbb{R}\left\{\zeta_o^{(m)}\right\}$) and $y_i^{(p)} = \mathbb{I}\left\{\zeta_i^{(p)}\right\}$ ($y_o^{(m)} = \mathbb{I}\left\{\zeta_o^{(m)}\right\}$), $p = 1, ..., P$ ($m = 1, ..., M$) are the coordinates of the $P$ ($M$) vertexes of the discretization of $\partial_i\Gamma$ ($\partial_o\Gamma$) [Fig. 1(*a*)].

The expressions (21) and (7) are finally substituted in (3) to define the overall mapping function $\boldsymbol{\tau}$ to be used in (8) and in (5)-(6) for synthesizing the *FMD* that exactly radiates the desired field.

It is worth remarking that (*i*) the proposed approach naturally handles arbitrarily-shaped lens with contours $\partial_i\Gamma$ and $\partial_o\Gamma$ modeled through their $P$/$M$-vertexes polygonal representations, $P$ and $M$ being user-defined parameters [Fig. 1(*c*)] and that (*ii*) no other control parameters have to be set, thus inherently avoiding possible misconfiguration/calibration issues.

## IV. Numerical Results and Analysis

This section has a twofold objective. On the one hand, it is aimed at illustrating the application of the proposed *SC-CTE* solution strategy to the design of advanced *FMD*s. On the other hand, it is devoted to assess the advantages and the potentialities of such a paradigm, also in comparison with state-of-the-art *TE* methods, when dealing with radiating systems featuring doubly-connected metalenses.

The first benchmark experiment is devoted to the synthesis of a *FMD* conformal to a square mast [Fig. 2(*c*)] that radiates the same field of an ideal circular layout [Fig. 2(*b*)]. More specifically, the *FMD* is composed by a $N = 65$ elements array arranged on a square contour $\partial_i\Gamma$ of side $10\,\lambda$ ($\nu_i = 5\sqrt{2}\,\lambda$) and coated with a metalens extended over a doubly-connected region $\Gamma$ whose inner boundary coincides with $\partial_i\Gamma$, while the external circular contour has a radius equal to $\nu_o = 14\,\lambda$ ["Square mast" - Fig. 2(*c*)]. The target field [Fig. 2(*a*)] is equal to that generated in free-space by a $N = 65$ isophorically-fed (i.e., $\{J_n^* = 1.0;\ n = 1, ..., N\}$) circular reference array [Fig. 2(*b*)].

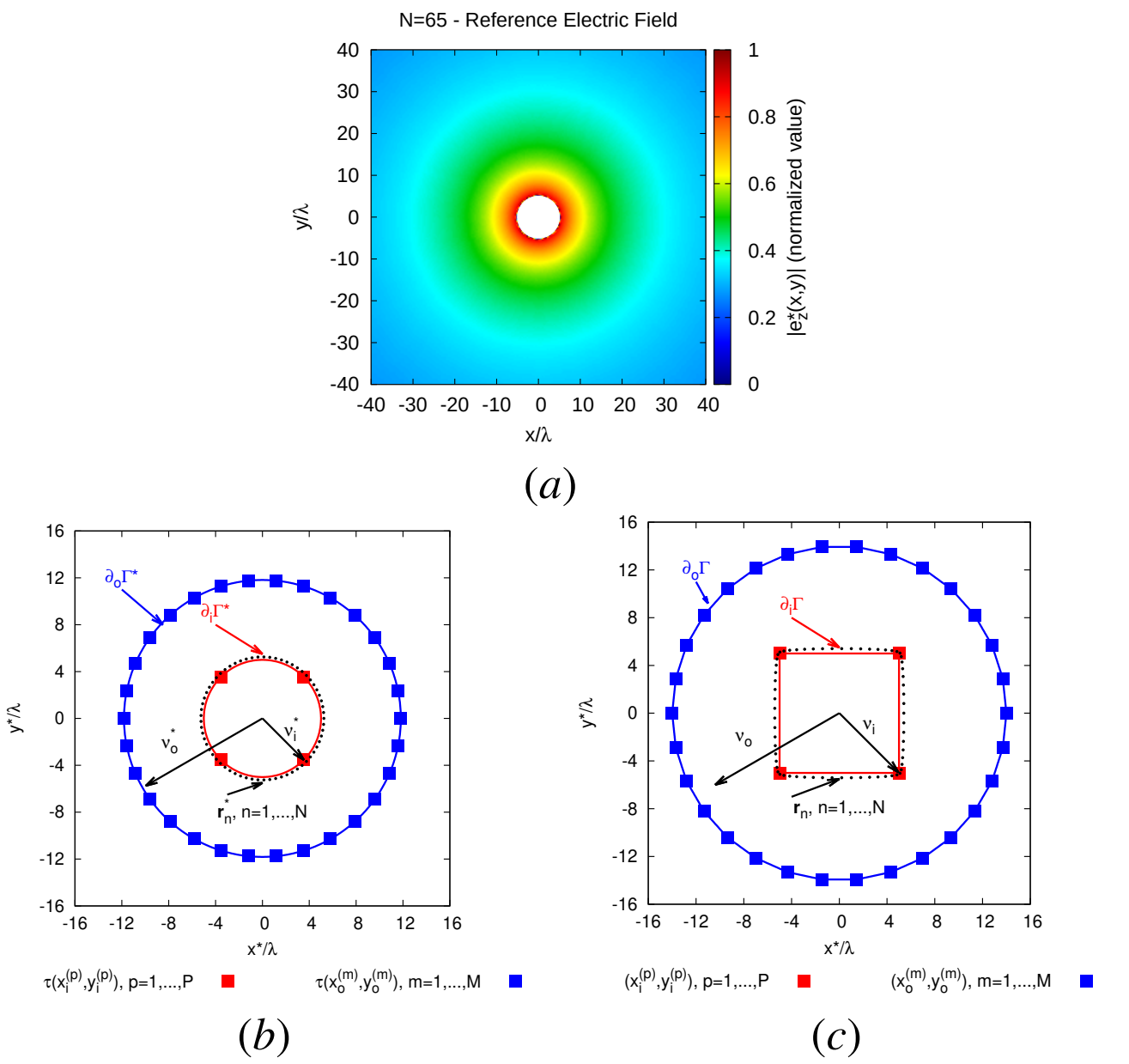

Figure 2. *Numerical Validation* ("*Square Mast*", $N = 65$, $P = 4$, $M = 30$) - Plot of (*a*) the target electric field distribution, $e_z^*(\mathbf{r})$, radiated by the (*b*) circular target geometry and geometry of (*c*) the *SC-CTE* synthesized *FMD*.

In order to apply the proposed *SC-CTE* strategy, the inner and the outer contours of the *FMD* (i.e., $\partial_i\Gamma$ and $\partial_o\Gamma$) have been uniformly sampled in $P = 4$, $\left\{\left(x_i^{(p)}, y_i^{(p)}\right);\ p = 1, ..., P\right\}$, and $M = 30$, $\left\{\left(x_o^{(m)}, y_o^{(m)}\right);\ m = 1, ..., M\right\}$, points with coordinates $y_{i/o}^{(p/m)} = \nu_{i/o}\cos\left[\frac{\pi}{4} + (P/M - 1)\times\frac{\pi}{2}\right]$ and $y_{i/o}^{(p/m)} = \nu_{i/o}\sin\left[\frac{\pi}{4} + (P/M - 1)\times\frac{\pi}{2}\right]$ [Fig. 2(*c*)]. Successively, the *SC Accessory Parameters* problem has been solved by minimizing (29) and yielding the values $C = 1.39\times 10^{1} + j6.63\times 10^{-9}$, $\mu = 4.23\times 10^{-1}$ [$\lambda$], and the prevertexes in Fig. 2(*b*). These latter values/coordinates have been then substituted in (21) to determine the desired mapping function $\boldsymbol{\tau}(\mathbf{r}^*)$, which has been finally used in (8) and in (5)-(6) to derive the lens dielectric properties and the resulting array configuration. For the sake of numerical efficiency and accuracy, the cylindrical components of the material tensors in (6) have been actually computed as in [32]

$$\overline{\overline{\varepsilon}}(\mathbf{r}) = \begin{bmatrix} \varepsilon_{\rho\rho}(\mathbf{r}) & \varepsilon_{\rho\varphi}(\mathbf{r}) & \varepsilon_{\rho z}(\mathbf{r}) \\ \varepsilon_{\varphi\rho}(\mathbf{r}) & \varepsilon_{\varphi\varphi}(\mathbf{r}) & \varepsilon_{\varphi z}(\mathbf{r}) \\ \varepsilon_{z\rho}(\mathbf{r}) & \varepsilon_{z\varphi}(\mathbf{r}) & \varepsilon_{zz}(\mathbf{r}) \end{bmatrix} = \left.\frac{\mathcal{J}_{cyl}[\boldsymbol{\tau}(\mathbf{r}^*)]\overline{\overline{\varepsilon}}(\mathbf{r}^*)\mathcal{J}_{cyl}^T[\boldsymbol{\tau}(\mathbf{r}^*)]}{\det\{\mathcal{J}_{cyl}[\boldsymbol{\tau}(\mathbf{r}^*)]\}}\right|_{\mathbf{r}^*=\mathbf{r}} \qquad (22)$$

where $\mathcal{J}_{cyl}[\boldsymbol{\tau}(\mathbf{r}^*)]$ is the Jacobian tensor in cylindrical coordinates given by

$$\mathcal{J}_{cyl}[\boldsymbol{\tau}(\mathbf{r}^*)] = \begin{bmatrix} \frac{\partial\tau_\rho(\mathbf{r}^*)}{\partial\rho^*} & \frac{1}{\rho^*}\frac{\partial\tau_\rho(\mathbf{r}^*)}{\partial\varphi^*} & 0 \\ \tau_\rho(\mathbf{r}^*)\frac{\partial\tau_\varphi(\mathbf{r}^*)}{\partial\rho^*} & \frac{\tau_\rho(\mathbf{r}^*)}{\rho^*}\frac{\partial\tau_\varphi(\mathbf{r}^*)}{\partial\varphi^*} & 0 \\ 0 & 0 & 1 \end{bmatrix}, \qquad (23)$$

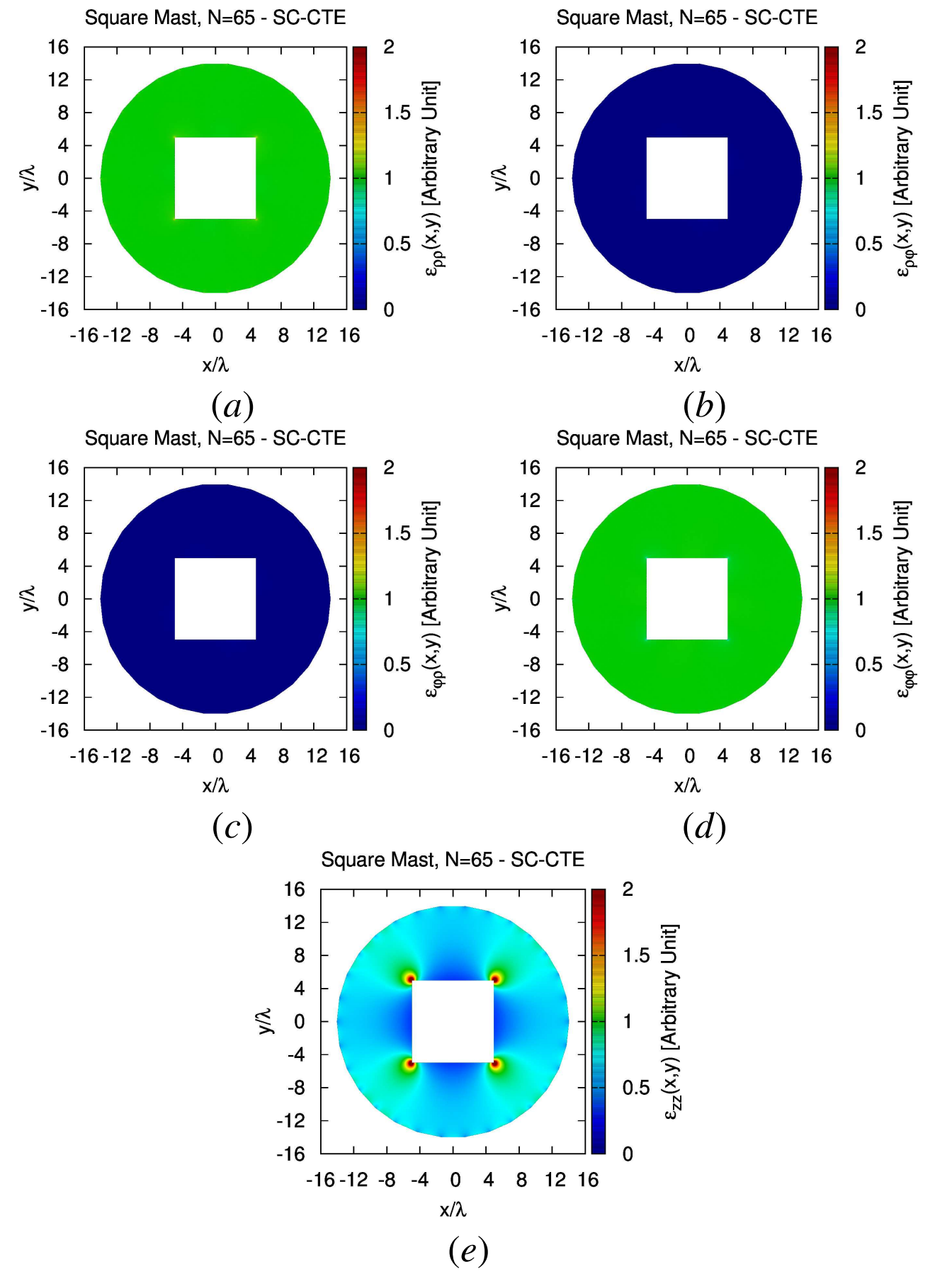

Figure 3. *Numerical Validation* ("*Square Mast*", $N = 65$, $P = 4$, $M = 30$) - Permittivity distributions of the *SC-CTE* lens layout: (*a*) $\varepsilon_{\rho\rho}(\mathbf{r})$, (*b*) $\varepsilon_{\rho\varphi}(\mathbf{r})$, (*c*) $\varepsilon_{\varphi\rho}(\mathbf{r})$, (*d*) $\varepsilon_{\varphi\varphi}(\mathbf{r})$, and (*e*) $\varepsilon_{zz}(\mathbf{r})$.

$\tau_\rho(\mathbf{r}^*) = \cos(\varphi^*)\tau_x(\mathbf{r}^*) + \sin(\varphi^*)\tau_y(\mathbf{r}^*)$ and $\tau_\varphi(\mathbf{r}^*) = -\sin(\varphi^*)\tau_x(\mathbf{r}^*) + \cos(\varphi^*)\tau_y(\mathbf{r}^*)$ being the cylindrical components of the *SC-CTE* transformation. Analogous expressions hold true for $\overline{\overline{\mu}}(\mathbf{r})$. It is worth pointing out that the cylindrical representation significantly simplifies the calculations since, by substituting (23) in (22), it turns out that $\varepsilon_{\rho z}(\mathbf{r}) = \varepsilon_{\varphi z}(\mathbf{r}) = \varepsilon_{z\rho}(\mathbf{r}) = \varepsilon_{z\varphi}(\mathbf{r}) = 0$, regardless of the geometries at hand.

Figure 3 shows the plots of the relative permittivity distributions within the lens support, $\Gamma$, the permeability maps being equal for symmetry reasons. As it can be observed, the off-diagonal elements of $\overline{\overline{\varepsilon}}(\mathbf{r})$ are exactly zero [Figs. 3(*b*)-3(*c*)], while the values of the $\rho$-th and $\varphi$-th diagonal entries are equal to the free-space permittivities [i.e., $\varepsilon_{\rho\rho}(\mathbf{r}) = \varepsilon_{\varphi\varphi}(\mathbf{r}) = 1.0$ - Fig. 3(*a*) and Fig. 3(*d*)]. In the overall, the synthesized material turns out to be uniaxially anisotropic, with optic axis along to the $\widehat{z}$ direction, unlike the metalenses yielded from advanced *QCTE* strategies that only approximate uniaxiality [7]. As for the $\varepsilon_{zz}(\mathbf{r})$ distribution, relatively stronger inhomogeneities arise in correspondence with the mast corners [Fig. 3(*e*)], as actually expected from *TE* theory, since sharp corners in the transformed regions are generally associated to permittivity peaks [7]. However, they are not so high [i.e., $\varepsilon_{zz}(\mathbf{r}) \leq 2.0$ - Fig. 3(*e*)] thanks to the adopted formulation, which relies on the availability of the explicit *SC* formula (16) for conformal mapping [24][25][30].

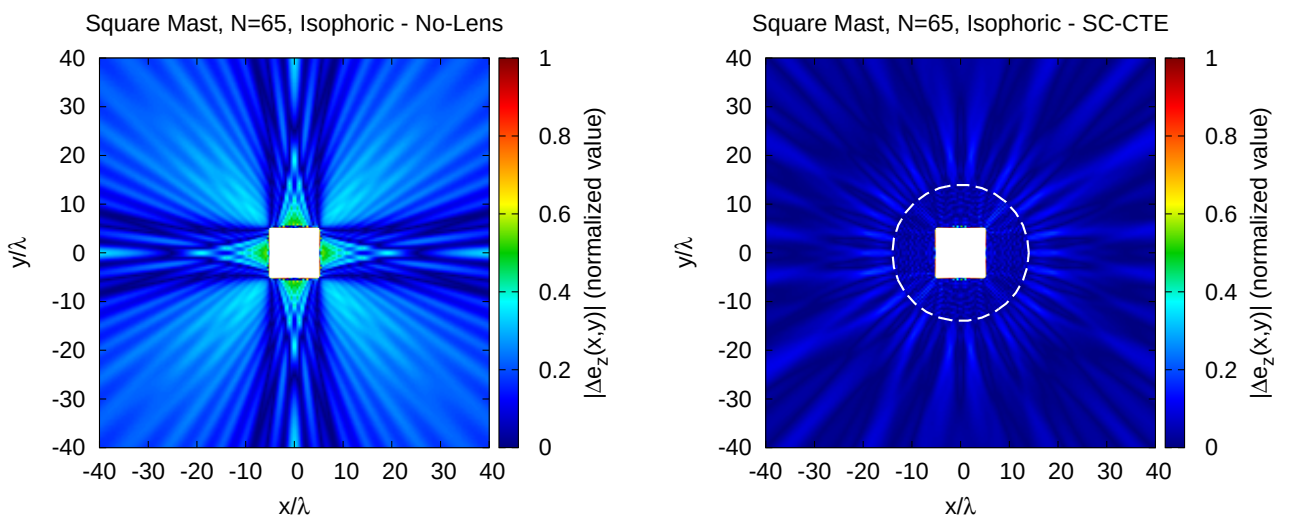

(*a*) (*b*)

Figure 4. *Numerical Validation* ("*Square Mast*", $N = 65$, $P = 4$, $M = 30$) - Plot of the magnitude of the normalized difference,$\left|\Delta e_z\left(\mathbf{r}\right)\right|$, when the array radiates (*a*) in free-space (i.e., without the lens) and (*b*) with the *SC-CTE* anisotropic lens.

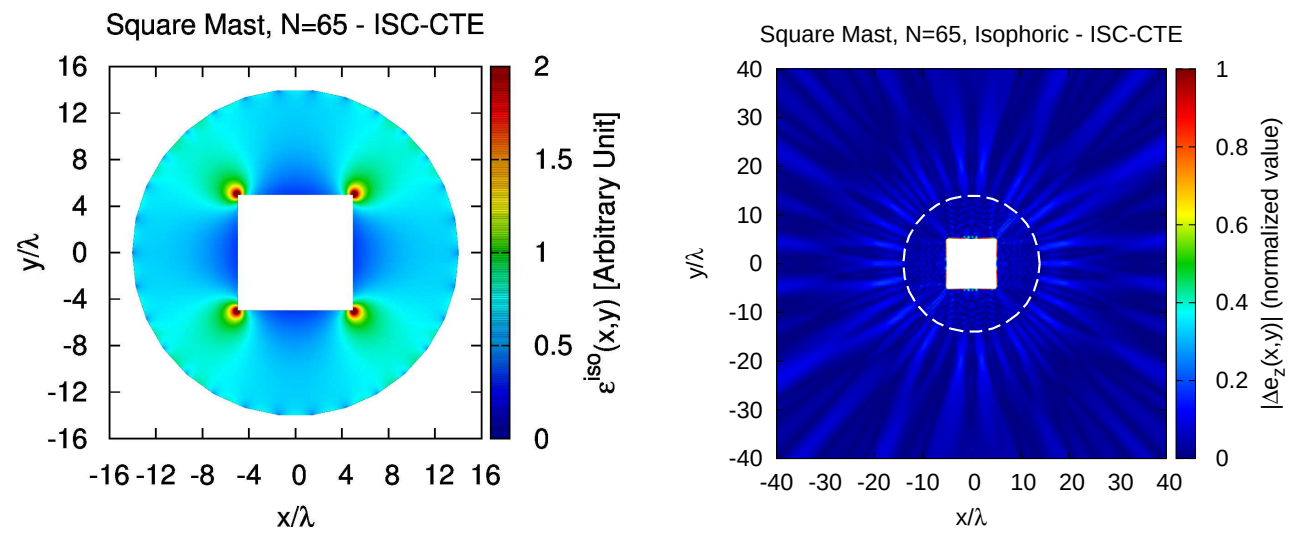

(*a*) (*b*)

Figure 5. *Numerical Validation* ("*Square Mast*", $N = 65$, $P = 4$, $M = 30$) - Permittivity distribution of (*a*) the isotropic simplification of the *SC-CTE* metalens, $\varepsilon^{iso}\left(\mathbf{r}\right)$, and plot of (*b*) the corresponding normalized field difference, $\left|\Delta e_z\left(\mathbf{r}\right)\right|$.

Concerning the matching of the *SC-CTE* field and the target one, the plot and the values of the magnitude of the normalized difference $\Delta e_z$ [7]

$$\Delta e_z\left(\mathbf{r}\right) \triangleq \frac{e_z\left(\mathbf{r}\right) - e_z^*\left(\mathbf{r}\right)}{\max_{\mathbf{r}\in\Gamma}\left|e_z^*\left(\mathbf{r}\right)\right|} \tag{24}$$

with [Fig. 4(*b*)] and without the lens [Fig. 4(*a*)] assess the efficiency of the *SC-CTE* device [i.e., $\left|\Delta e_z\left(\mathbf{r}\right)\right| \leq 1.7 \times 10^{-1}$ - Fig. 4(*b*)] and the performance improvement with respect to the array without the lens [i.e., $\left|\Delta e_z\left(\mathbf{r}\right)\right| \leq 5.7 \times 10^{-1}$ - Fig. 4(*a*)].

Although such a result points out the effectiveness of the *SC-CTE* in designing doubly-connected metalenses, the synthesized *FMD* features uniaxially anisotropic materials (Fig. 3) whose implementation may be non-trivial. Thus, it is of practical interest to check the possibility of achieving similar performance with an isotropic approximation of the lens material to enable simpler fabrication procedures [10]. Towards this end and owing to the transverse-magnetic (*TM*) nature of the propagation problem at hand [7], the isotropic and non-magnetic (i.e., its permeability being set to unity $\forall\mathbf{r} \in \Gamma$) approximation of (22), hereinafter referred to as *isotropic SC-CTE* (*ISC-CTE*), is derived $\overline{\overline{\varepsilon}}\left(\mathbf{r}\right) = \varepsilon^{iso}\left(\mathbf{r}\right)\overline{\overline{I}}$, $\overline{\overline{I}}$ being the unitary tensor [$\varepsilon^{iso}\left(\mathbf{r}\right)$ - Fig. 5(*a*)], starting from the computed *SC-CTE* mapping function, as follows [28]

$$\begin{cases} \varepsilon^{iso}\left(\mathbf{r}\right) \triangleq \left(\det\left\{\mathcal{J}_{cyl}\left[\boldsymbol{\tau}\left(\mathbf{r}^*\right)\right]\right\}\right)^{-1}\Big|_{\mathbf{r}^*=\mathbf{r}} & \text{if } \mathbf{r} \in \Gamma. \\ \mu^{iso}\left(\mathbf{r}\right) \triangleq 1.0 \end{cases} \tag{25}$$

The result is very promising in this case, as well, as confirmed by the distribution of the normalized difference field in Fig. 5(*b*), whose maximum value does not exceed the threshold of $\left|\Delta e_z\left(\mathbf{r}\right)\right| \leq 2.0 \times 10^{-1}$, which is slightly greater than that for the full anisotropic case [Fig. 5(*b*) vs. Fig. 4(*b*)]. Such an outcome highlights a key advantage of the *SC-CTE* technique with respect to state-of-the-art *QCTE* numerical methods, since the isotropic approximations of the synthesized *FMD*s behave almost identically [Fig. 5(*b*) vs. Fig. 4(*b*)] thanks to the *conformal* nature of the *SC* mapping formula.

The next experiment deals with the same mast structure [Fig. 2(*c*)], but only the $N_1 = 16$ ($N_1 < N$) array elements, placed as shown in the inset of Fig. 6(*a*), are isophorically-excited to radiate a beam along broadside (i.e., $\varphi_0 = 90$ [deg]). Since the *SC-CTE* technique builds upon the invariance of Maxwell's equations under coordinate mapping [2], the *FMD* must actually afford equivalent performance regardless of the excitations set, $\{J_n^*;\, n = 1, ..., N\}$. Therefore, the lens profile in Fig. 3 has been kept unaltered. As expected, such a new *FMD* (the array excitations being different) behaves like the previous one, with all $N$ elements turned on, in matching the target field ($\left|\Delta e_z\left(\mathbf{r}\right)\right|_{N_1} \approx \left|\Delta e_z\left(\mathbf{r}\right)\right|_{N}$) being $\left|\Delta e_z\left(\mathbf{r}\right)\right|_{N_1} \leq 2.7 \times 10^{-1}$, while the radiation without the lens is very far from the objective [Fig. 6(*a*) and Fig. 6(*b*)]. It is also worth noticing that the non-magnetic and isotropic approximation (25) of the *SC-CTE* lens gives almost identical field manipulation capabilities of its uniaxial counterpart [Fig. 6(*d*) vs. Fig. 6(*c*)], despite the lens-material simplification. Those outcomes are further pointed out by the behavior of the normalized far-field power patterns, $\mathcal{P}\left(\varphi\right)$, in Fig. 6(*a*) and the corresponding far-field matching errors ($\eta \triangleq \frac{\int_{-\pi}^{\pi}\left|\mathcal{P}(\varphi)-\mathcal{P}^*(\varphi)\right|d\varphi}{\int_{-\pi}^{\pi}\mathcal{P}^*(\varphi)d\varphi}$, $\mathcal{P}^*\left(\varphi\right)$ being the normalized far-field power patterns of the target field) as well as the values of the absolute (i.e., non-normalized) pattern features in Tab. I.

Table I
*Numerical Validation* ("*Square Mast*", $N = 65$) - PATTERN FEATURES AND MATCHING INDEXES.

| | *No-Lens* | *SC-CTE* | *ISC-CTE* |
|---|---|---|---|
| $\eta$ [%] | 73.4 | 1.31 | 1.36 |
| *SLL* [dB] | $-0.18$ | $-10.67$ | $-10.44$ |
| $D_{\max}$ | 5.24 | 13.38 | 13.36 |
| *HPBW* [deg] | 23.23 | 6.78 | 6.81 |

The dependence of the field manipulation capabilities on the variations of the array excitations, while keeping the same array architecture with $N_1 = 16$ active elements, is carried out next (Fig. 7). The plots of $\mathcal{P}\left(\varphi\right)$ and $\mathcal{P}^*\left(\varphi\right)$ when steering the beam within the range $\varphi_0 \in [90, 120]$ [deg] show that the patterns radiated by the *SC-CTE FMD* [Fig. 7(*a*)] as well as from its isotropic version (25) [Fig. 7(*b*)] faithfully match the target ones. For completeness, the behaviour of the far-field figures of merit is shown in Fig. 8. As it can be observed, the deviations from the target values are very limited being $\eta^{SC-CTE}\left(\varphi_0\right) \leq 6\%$ and $\eta^{ISC-CTE}\left(\varphi_0\right) \leq 7\%$ [Fig. 8(*a*)], $\left|D_{\max}^{SC-CTE}\left(\varphi_0\right) - D_{\max}^*\left(\varphi_0\right)\right| \leq 0.21$ [dB] and $\left|D_{\max}^{ISC-CTE}\left(\varphi_0\right) - D_{\max}^*\left(\varphi_0\right)\right| \leq 0.32$ [dB] [Fig. 8(*b*)],

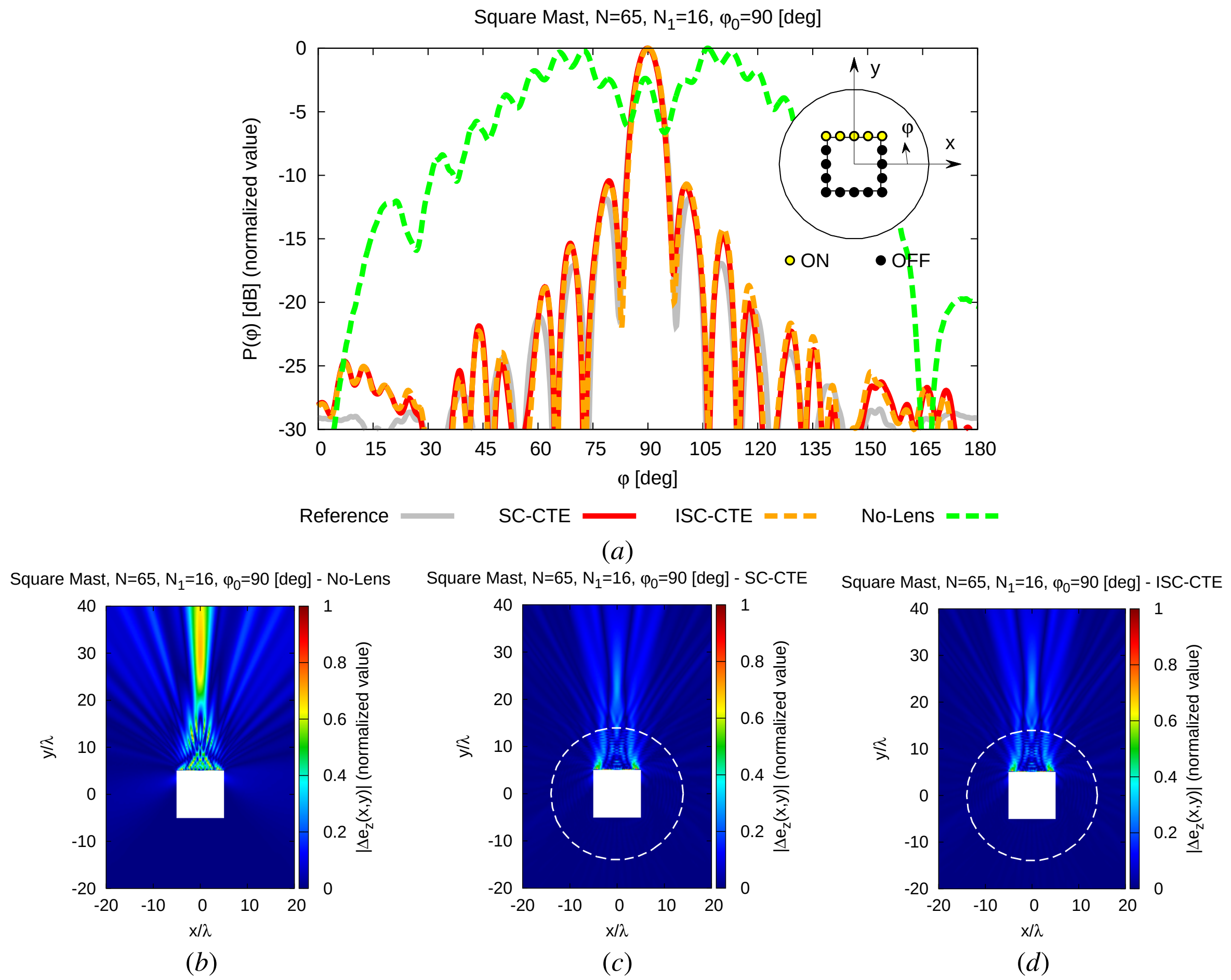

Figure 6. *Numerical Validation* ("*Square Mast*", $N = 65$, $P = 4$, $M = 30$ - $N_1 = 16$, $\varphi_0 = 90$ [deg]) - Plot of ($a$) the normalized power pattern, $\mathcal{P}(\varphi)$, of the target/reference array along with those with/without the *SC-CTE* and the *ISC-CTE* metalenses. Map of the magnitude of the normalized difference field,$|\Delta e_z(\mathbf{r})|$, when the array radiates ($b$) in free-space and in the presence of ($c$) the anisotropic and ($d$) the isotropically-simplified *SC-CTE* metalenses.

$\left|HPBW^{SC-CTE}(\varphi_0) - HPBW^*(\varphi_0)\right| \leq 0.43$ [deg] and $\left|HPBW^{ISC-CTE}(\varphi_0) - HPBW^*(\varphi_0)\right| \leq 0.46$ [deg], and $\left|SLL^{SC-CTE}(\varphi_0) - SLL^*(\varphi_0)\right| \leq 1.15$ [dB] and $\left|SLL^{ISC-CTE}(\varphi_0) - SLL^*(\varphi_0)\right| \leq 1.75$ [dB] [Fig. 8($c$)], despite the geometrical differences between the reference circular array arrangement and the square actual one [Fig. 2($b$) vs. Fig. 2($c$)].

The third numerical experiment is aimed at comparing the proposed *SC-CTE* paradigm with recently introduced *quasi-conformal* TE methods [1][7]. Towards this end, the *generalized QCTE* technique [7] is considered hereinafter because its flexibility, effectiveness, and reliability in approximating conformal mappings as proven in different *TE* applicative scenarios [7][19][21]. However, since *QCTE* methods cannot natively handle doubly-connected regions such as those of interest (Fig. 1) [7], the problem at hand has been addressed by combining two "half-space" lenses analogous to those designed in [7]. Figure 9($a$) shows the plots of $\mathcal{P}(\varphi)$ for the target, the *SC-CTE*, and the *QCTE* layouts when the $N_1 = 16$ array elements in the inset of Fig. 9($a$) are excited to steer the main beam along to $\varphi_0 = 100$ [deg]. As expected, the *SC*-based *FMD* outperforms the *QCTE* one as indicated by the red curve versus the blue one in Fig. 9($a$)] and numerically confirmed by the corresponding far-field figures of merit [e.g., $\eta^{SC-CTE} \approx 2.01\%$ vs. $\eta^{QCTE} \approx 2.74\%$; $\left|SLL^{SC-CTE} - SLL^*\right| \approx 1.26$ [dB] vs. $\left|SLL^{QCTE} - SLL^*\right| \approx 3.15$ [dB] - Fig. 9($a$)]. The improvement of the *SC-CTE* becomes impressive when analyzing the isotropic realizations of the same lenses [i.e., orange vs. cyan lines - Fig. 9($a$)]. As a matter of fact, the *ISC-CTE FMD* works similarly to its anisotropic counterpart [e.g., $\eta^{ISC-CTE} \approx 2.24\%$ - Fig. 9($a$)] and, by the way, even better than the fully-anisotropic *QCTE*-based design [e.g., $\left|SLL^{ISC-CTE} - SLL^*\right| \approx 1.30$ [dB] - Fig. 9($a$)]. On the contrary, the isotropic version of the *QCTE* device is not able to control the sidelobe profile with a complete loss of the main beam focusing [i.e., cyan line - Fig. 9($a$)] resulting in a very poor far-field matching [e.g., $\eta^{IQCTE} \approx 105\%$ - Fig. 9($a$)]. Such results are not surprising and they are theoretically expected because of the quasi-conformal (i.e., not conformal) nature of the *QCTE* formulation [7][19][21], unlike the the proposed *SC-CTE* technique. Consequently, the *QCTE* lens turns out considerably more anisotropic than the *SC-CTE* one and an isotropic simplification often affords poor performance especially when the array is fed with non-isophoric excitations [7]. To give some insights on such an item, the degree of

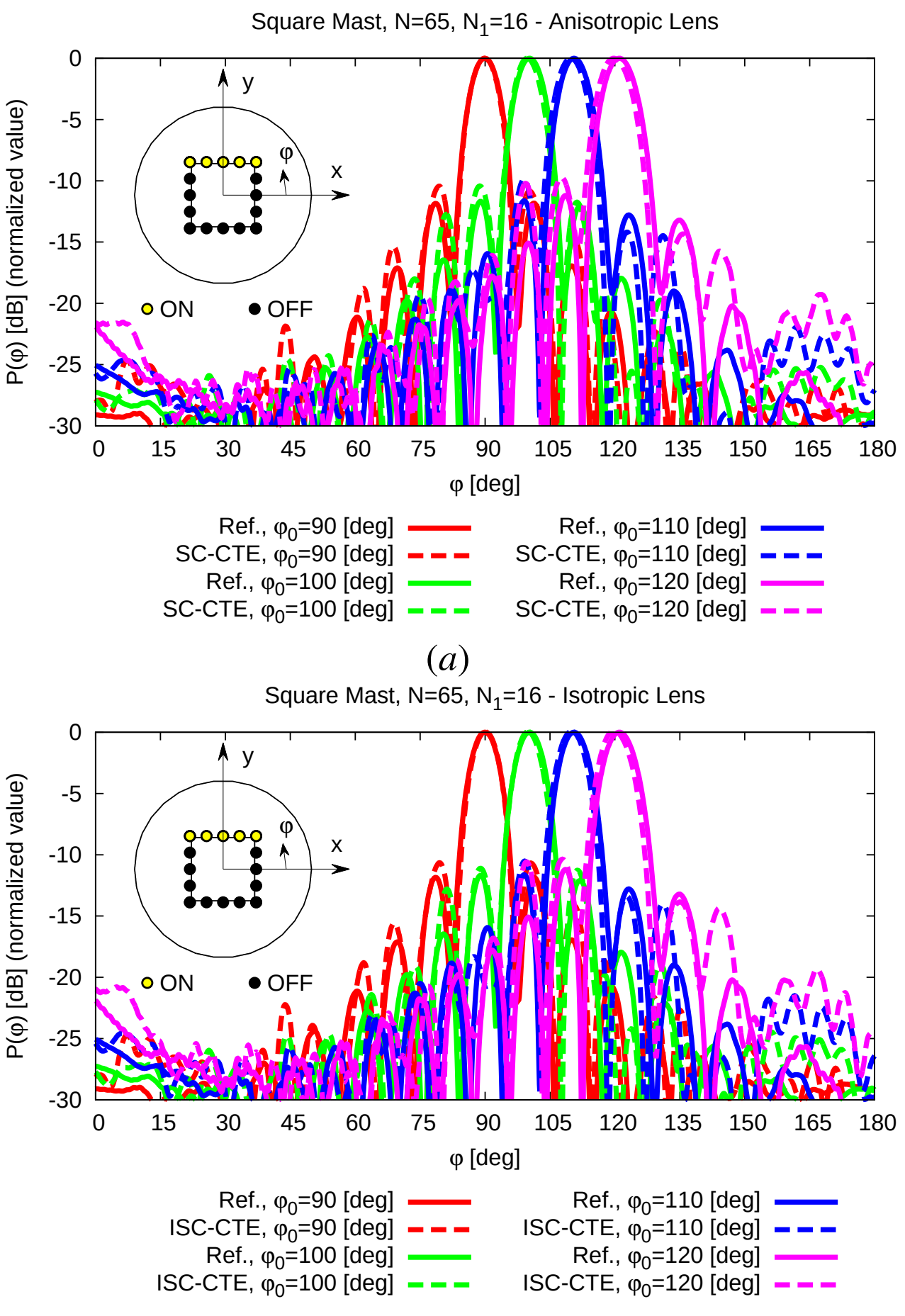

Figure 7. *Numerical Validation* ("*Square Mast*", $N = 65$, $P = 4$, $M = 30$ - $N_1 = 16$, $\varphi_0 \in [90, 120]$ [deg]) - Plot of the normalized power pattern, $\mathcal{P}(\varphi)$, radiated by the reference array and by *FMD* with (*a*) a *SC-CTE* or (*b*) as *ISC-CTE* lens.

anisotropy of the synthesized metalenses has been quantified by computing the *average fractional anisotropy* [21]

$$\alpha_F \triangleq \frac{1}{area(\Gamma)} \int_{\mathbf{r}\in\Gamma} \sqrt{\frac{3 \times \sum_{i=1}^{3} [\sigma_i(\mathbf{r}) - \sigma_{ave}(\mathbf{r})]^2}{2 \times \sum_{i=1}^{3} [\sigma_i(\mathbf{r})]^2}} d\mathbf{r} \tag{26}$$

and the *average relative anisotropy* [21]

$$\alpha_R \triangleq \frac{1}{area(\Gamma)} \int_{\mathbf{r}\in\Gamma} \sqrt{\frac{\sum_{i=1}^{3} [\sigma_i(\mathbf{r}) - \sigma_{ave}(\mathbf{r})]^2}{3 \times \sigma_{ave}(\mathbf{r})}} d\mathbf{r} \tag{27}$$

where $\sigma_{ave}(\mathbf{r}) \triangleq \frac{\sum_{i=1}^{3}\sigma_i(\mathbf{r})}{3}$ and $\sigma_i(\mathbf{r})$, $i = 1, ..., 3$, are the eigenvalues of $\overline{\overline{\varepsilon}}(\mathbf{r})$. The values reported in Tab. II confirm the reduction of the anisotropy when exploiting the *SC-CTE* approach instead of the state-of-the-art *QCTE* one (i.e., $\frac{\alpha_F\rfloor^{SC-CTE} - \alpha_F\rfloor^{QCTE}}{\alpha_F\rfloor^{QCTE}} \approx -29\%$ and $\frac{\alpha_R\rfloor^{SC-CTE} - \alpha_R\rfloor^{QCTE}}{\alpha_R\rfloor^{QCTE}} \approx -35\%$).
The advantages of using an *SC-CTE FMD* are even more remarkable when the beam is focused towards the lens regions with sharp permittivity variations, which usually yield to severe pattern distortions in state-of-the-art *QCTE* devices [7]. To check such a critical condition, the test case with the beam steered along $\varphi_0 = 145$ [deg] and the $N_2$ ($N_2 = 16$) corner

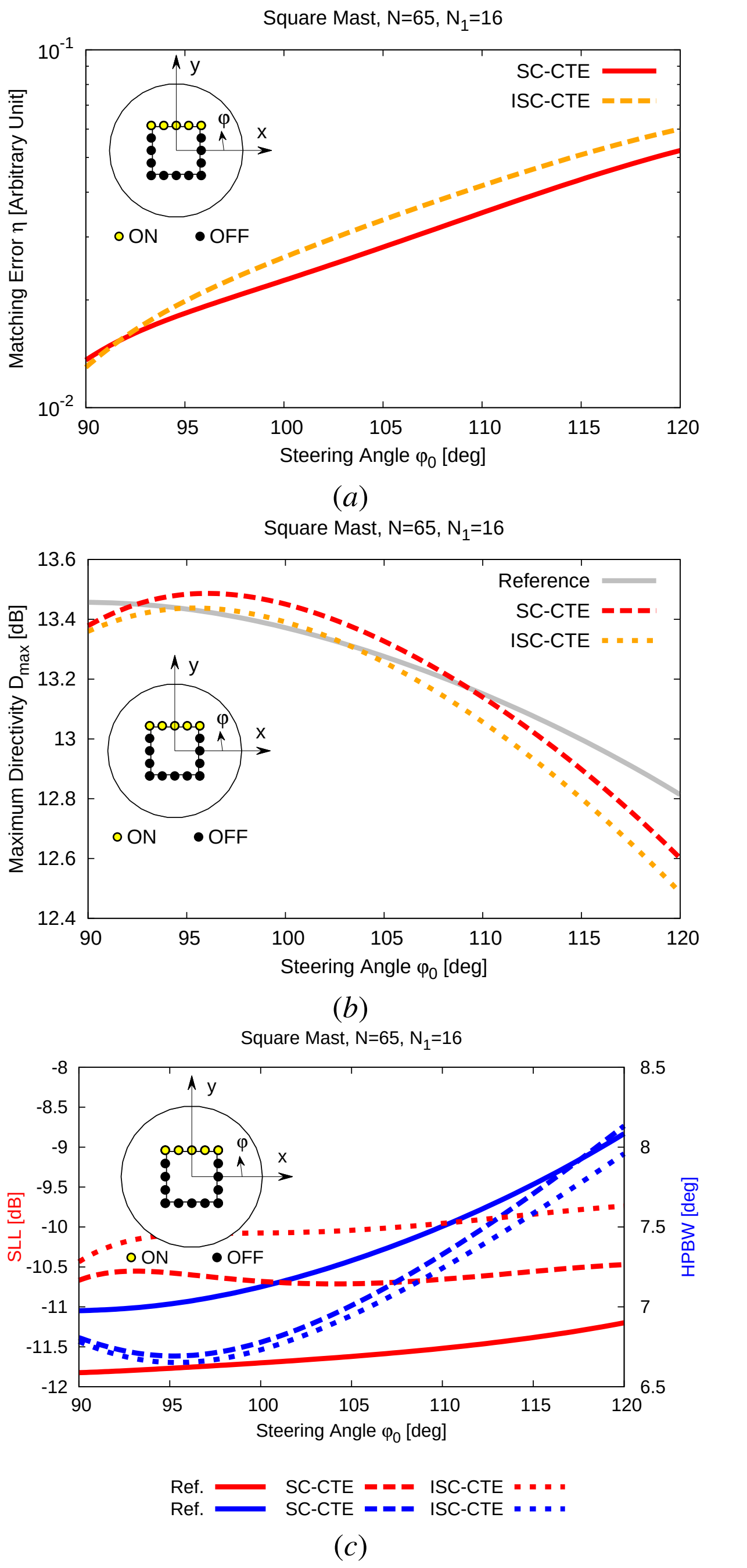

Figure 8. *Numerical Validation* ("*Square Mast*", $N = 65$, $P = 4$, $M = 30$, $N_1 = 16$, $\varphi_0 \in [90, 120]$ [deg]) - Plot of (*a*) $\eta$, (*b*) $D_{\max}$, and (*c*) the $SLL$ and $HPBW$ versus the steering angle value, $\varphi_0$.

elements turned on has been addressed [Fig. 9(*b*)]. Unlike the previous test case, the *QCTE* layout poorly matches the target pattern even when exploiting the fully-anisotropic lens [e.g., $\eta^{QCTE} \approx 69.42\%$ - Fig. 9(*b*)]. On the contrary, the *SC*-based implementation performs well in both the anisotropic [e.g., $\eta^{SC-CTE} \approx 2.99\%$ - Fig. 9(*b*)] and the isotropic [e.g., $\eta^{ISC-CTE} \approx 3.83\%$ vs. $\eta^{ISC-QCTE} \approx 80.31\%$ - Fig. 9(*b*)] cases. To give an idea of the matching/mismatching in the near-field region, as well, Figure 10 presents a pictorial representation of the field afforded by the different *FMD*s in terms of normalized difference values, $\Delta e_z$ [Figs. 10(*b*)-

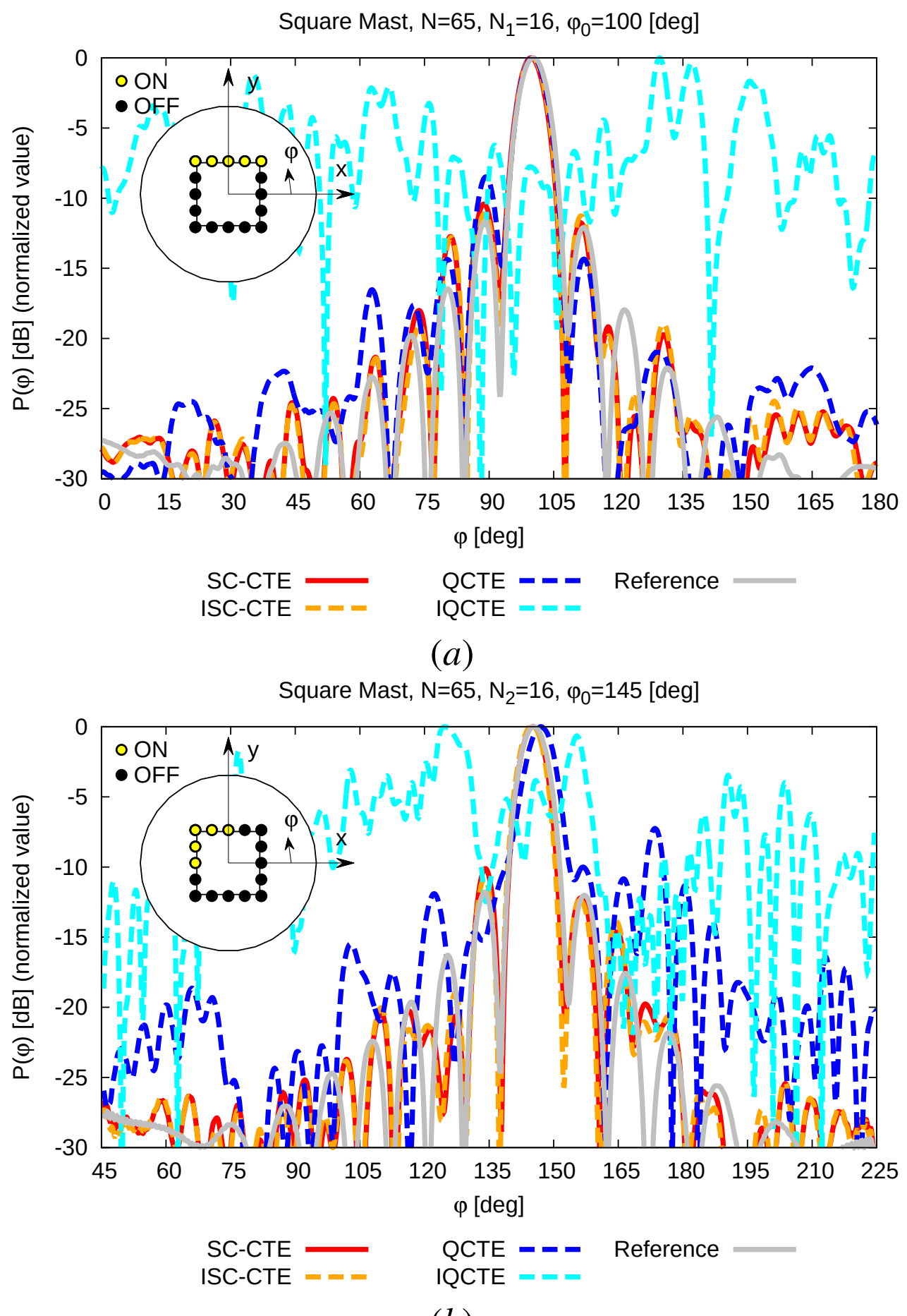

Figure 9. *Numerical Validation* ("*Square Mast*", $N = 65$, $P = 4$, $M = 30$) - Plot of the normalized power pattern, $\mathcal{P}(\varphi)$, of the reference array and of the *FMD*s with the *SC-CTE* and the *QCTE* lenses when exciting (*a*) the top ($N_1 = 16$) array elements to steer the beam along $\varphi_0 = 100$ [deg] and (*b*) the corner ($N_2 = 16$) array elements to steer the beam towards $\varphi_0 = 145$ [deg].

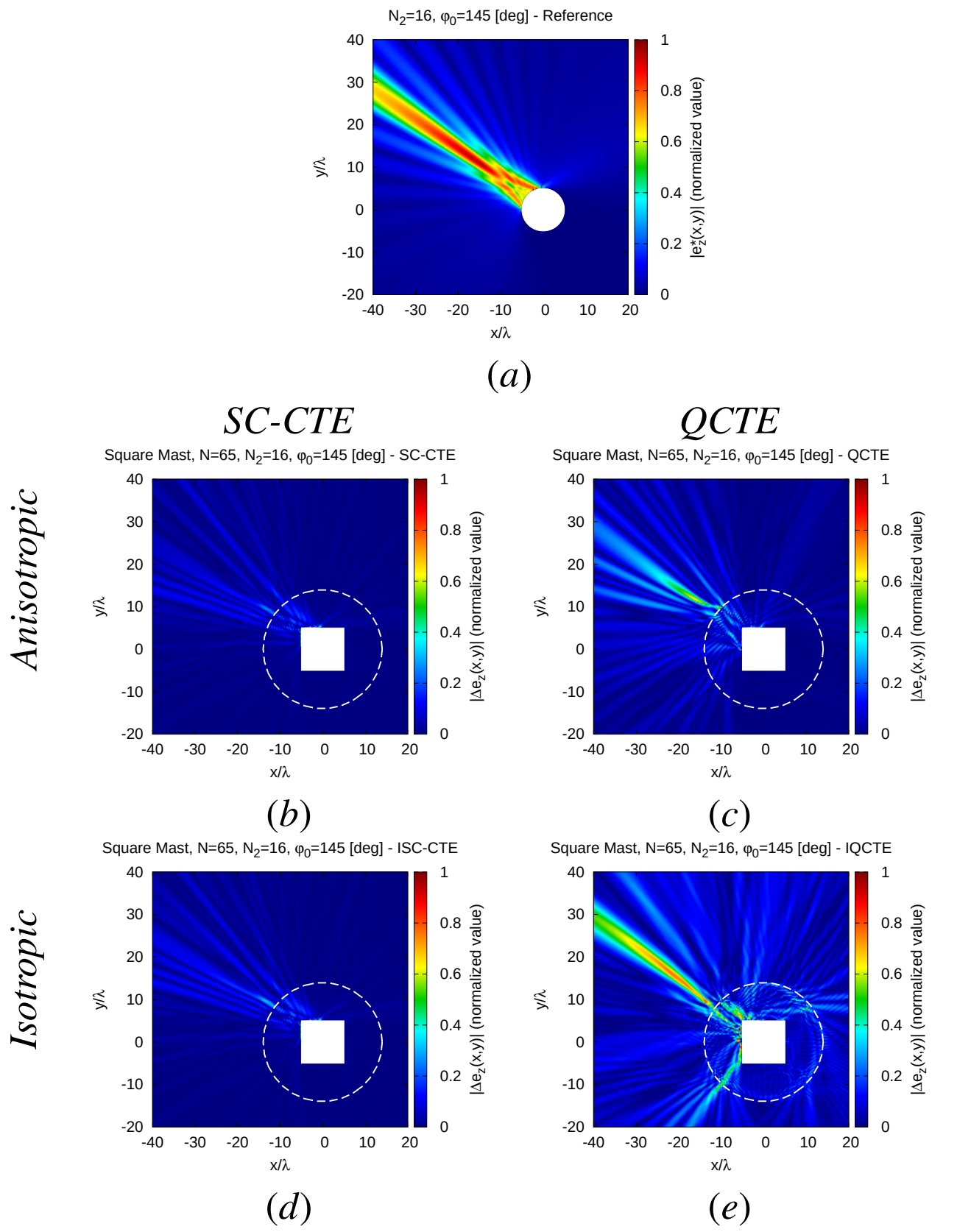

Figure 10. *Numerical Validation* ("*Square Mast*", $N = 65$, $P = 4$, $M = 30$ - $N_2 = 16$, $\varphi_0 = 145$ [deg]) - Plot of the magnitude of (*a*) the target field distribution, $e^*(\mathbf{r})$, and of the normalized difference field,$|\Delta e_z(\mathbf{r})|$, for the (*b*)(*d*) *SC-CTE* and (*c*)(*e*) *QCTE FMD*s with (*b*)(*c*) anisotropic and (*d*)(*e*) isotropic metalenses.

10(*e*)], with respect to the target distribution [Fig. 10(*a*)].

By extending the comparative analysis between the *SC-CTE* and the *QCTE* to different steering angles, still using the "corner element" setup in the inset of Fig. 11, it turns out that the performance of the *SC-CTE* device does not significantly degrade (e.g., $\eta^{SC-CTE}_{\varphi_0=135\,[deg]} \approx 2.35\%$ vs. $\eta^{SC-CTE}_{\varphi_0=165\,[deg]} \approx 10.76\%$ - Fig. 11) and the corresponding isotropic approximation (25) yields similar results (e.g., $\left|\eta^{ISC-CTE}(\varphi_0) - \eta^{SC-CTE}(\varphi_0)\right|_{\varphi_0\in[135,165]\,[deg]} \leq 1.4\%$ - Fig. 11). On the contrary, the *QCTE FMD* behaves analogously to the array without the lens regardless of $\varphi_0$ (e.g., $\eta^{QCTE}_{\varphi_0=165\,[deg]} \approx 94.61\%$ vs. $\eta^{No-Lens}_{\varphi_0=165\,[deg]} \approx 106.43\%$ - Fig. 11).

Moving from the "square mast" configuration [Fig. 2(*c*)] to a more complex scenario, the next experiment is concerned with an array of $N = 65$ elements uniformly-distributed on an hexagonal contour $\partial_i\Gamma$ and it deals with the synthesis of a *FMD*, which integrates a lens with a user-defined doubly-connected shape $\Gamma$ having an external circular contour $\partial_o\Gamma$ (i.e., "Hexagonal mast" - Fig. 12), that radiates the target field $e^*(\mathbf{r})$ in Fig. 2(*a*). Concerning the dielectric properties of the *SC-CTE* lens, $\overline{\overline{\varepsilon}}(\mathbf{r})$ [Figs. 12(*a*)-12(*e*)], and its isotropic approximation, $\varepsilon^{iso}(\mathbf{r})$ [Fig. 12(*f*)], the same conclusions drawn from the "square mast" case hold true since (*i*) the lens material presents a uniaxially anisotropy with $\widehat{z}$ optic axis [i.e., $\varepsilon_{\rho\varphi}(\mathbf{r}) = \varepsilon_{\varphi\rho}(\mathbf{r}) = 0.0$ - Fig. 12(*b*) and Fig. 12(*c*); $\varepsilon_{\rho\rho}(\mathbf{r}) = \varepsilon_{\varphi\varphi}(\mathbf{r}) = 1.0$ - Fig. 12(*a*) and Fig. 12(*d*)], (*ii*) the permittivity distribution has peaks at the sharp corners of the mast contour [Fig. 12(*e*)] with relatively limited magnitude [i.e., $\varepsilon_{zz}(\mathbf{r}) \leq 2.0$ - Fig. 12(*e*)] as a consequence of the *CTE* property of the *SC*-based method, (*iii*) the profile of the $\varepsilon_{zz}$-component is quite continuous without abrupt permittivity transitions, and (*iv*) the isotropic approximation yields a lens with dielectric properties close to $\varepsilon_{zz}(\mathbf{r})$ [Fig. 12(*f*) vs. Fig. 12(*e*)]. The analysis of the field distribution, $e(\mathbf{r})$, when all the $N = 65$ array elements are uniformly excited and they radiate in free space [Fig. 13(*a*)], in the presence of the *SC-CTE* lens [Fig. 13(*b*)] or the *ISC-CTE* one [Fig. 13(*c*)] highlights the effectiveness of the *SC FMD*s in fitting the user requirements being $\left|\Delta e_z^{SC-CTE}(\mathbf{r})\right| \leq 2.16 \times 10^{-1}$ [Fig. 13(*b*)] and $\left|\Delta e_z^{ISC-CTE}(\mathbf{r})\right| \leq 2.18 \times 10^{-1}$ [Fig. 13(*c*)], while unsatisfactory results are obtained without adding field-manipulating components [i.e., $\left|\Delta e_z^{No-Lens}(\mathbf{r})\right| \leq 5.73 \times 10^{-1}$ - Fig. 13(*a*)].

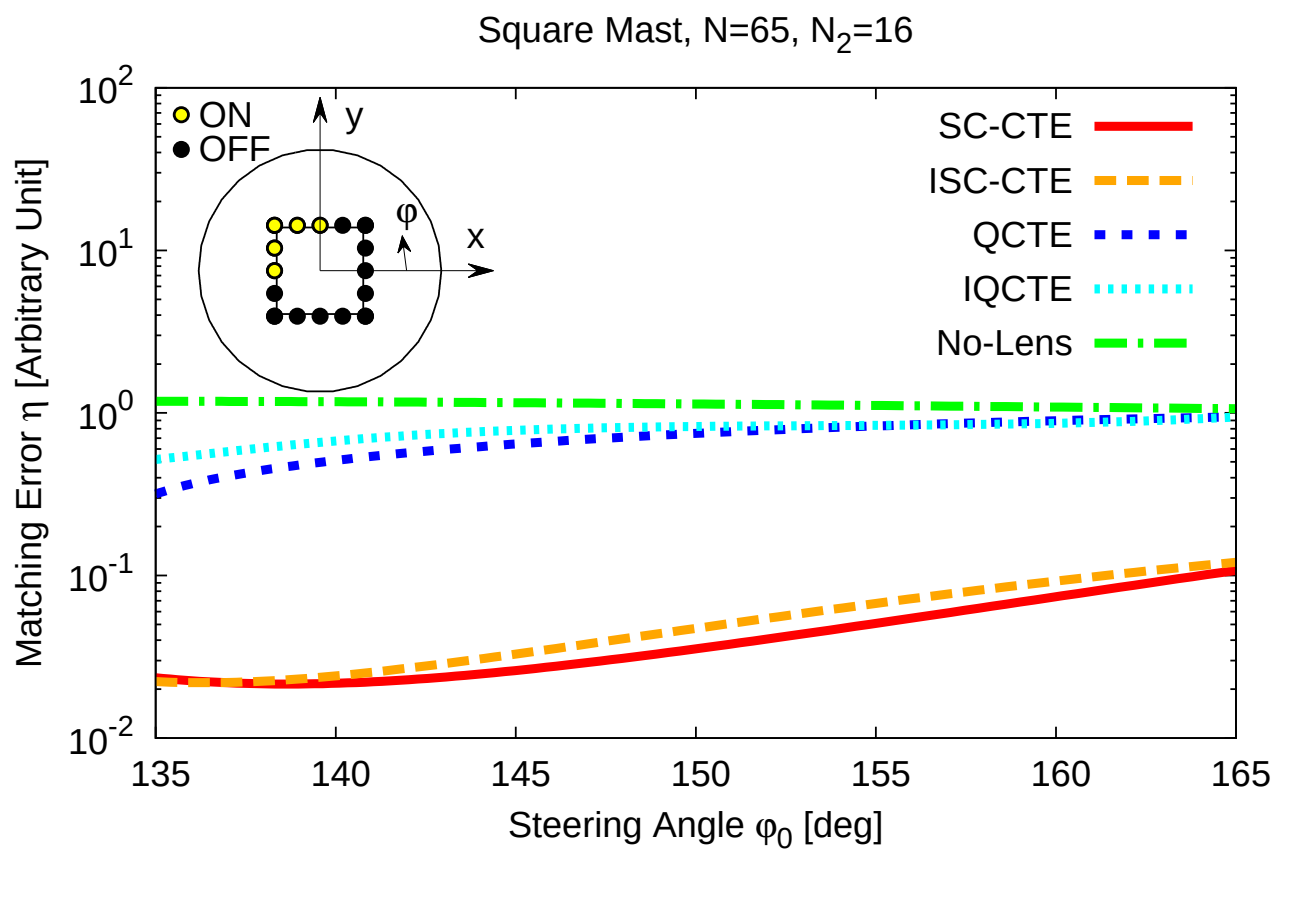

Figure 11. *Numerical Validation* ("*Square Mast*", $N = 65$, $P = 4$, $M = 30$ - $N_2 = 16$, $\varphi_0 \in [135, 165]$ [deg]) - Behavior of $\eta$ versus the steering angle, $\varphi_0$, when the array radiates in free-space and in the presence of anisotropic/isotropic *SC-CTE* and *QCTE* metalenses.

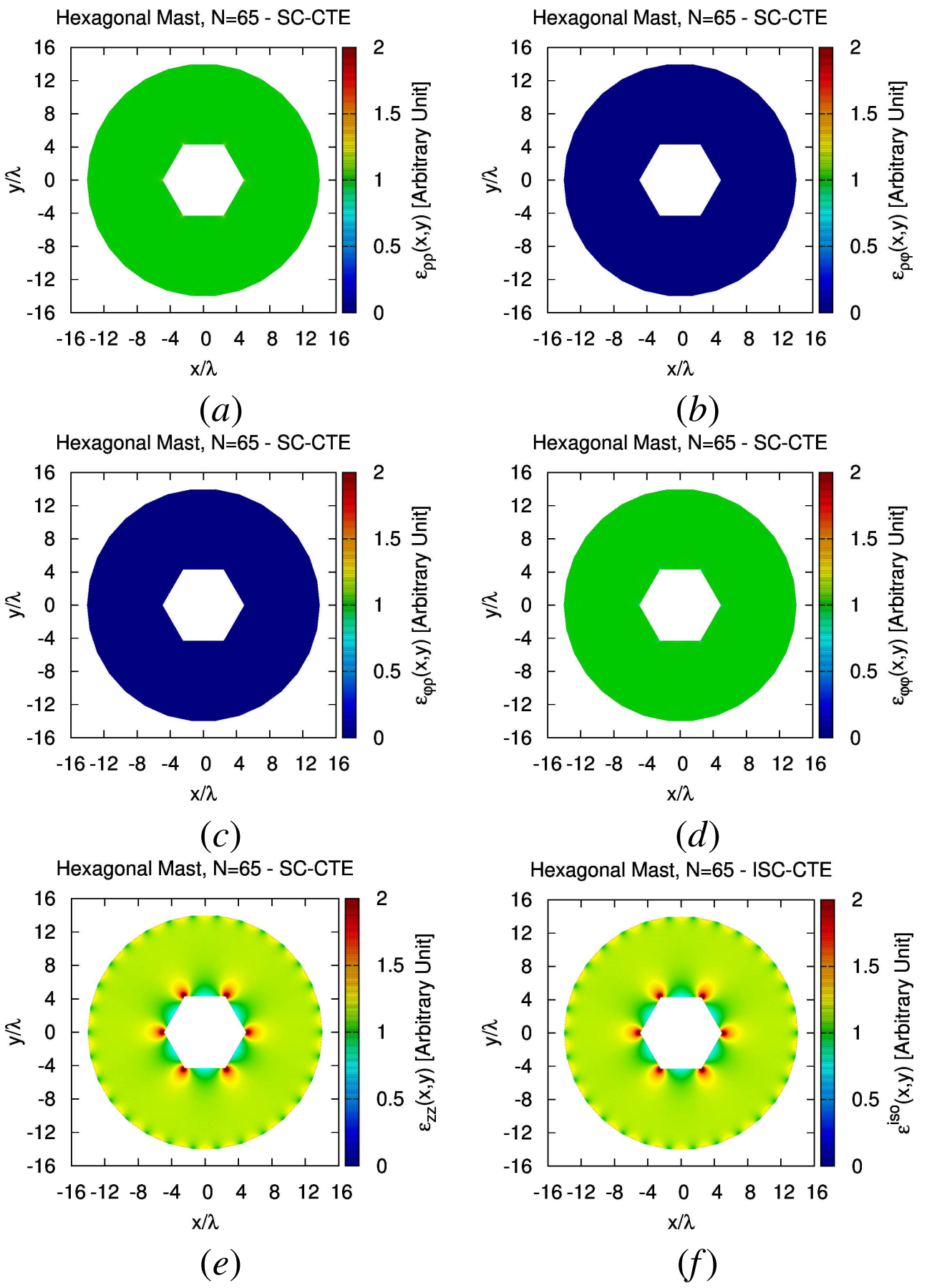

Figure 12. *Numerical Validation* ("*Hexagonal Mast*", $N = 65$, $P = 6$, $M = 30$) - Permittivity distributions of the *SC-CTE* lens [(*a*) $\varepsilon_{\rho\rho}(\mathbf{r})$, (*b*) $\varepsilon_{\rho\varphi}(\mathbf{r})$, (*c*) $\varepsilon_{\varphi\rho}(\mathbf{r})$, (*d*) $\varepsilon_{\varphi\varphi}(\mathbf{r})$, and (*e*) $\varepsilon_{zz}(\mathbf{r})$] and of (*f*) the *ISC-TE* lens, $\varepsilon^{iso}(\mathbf{r})$.

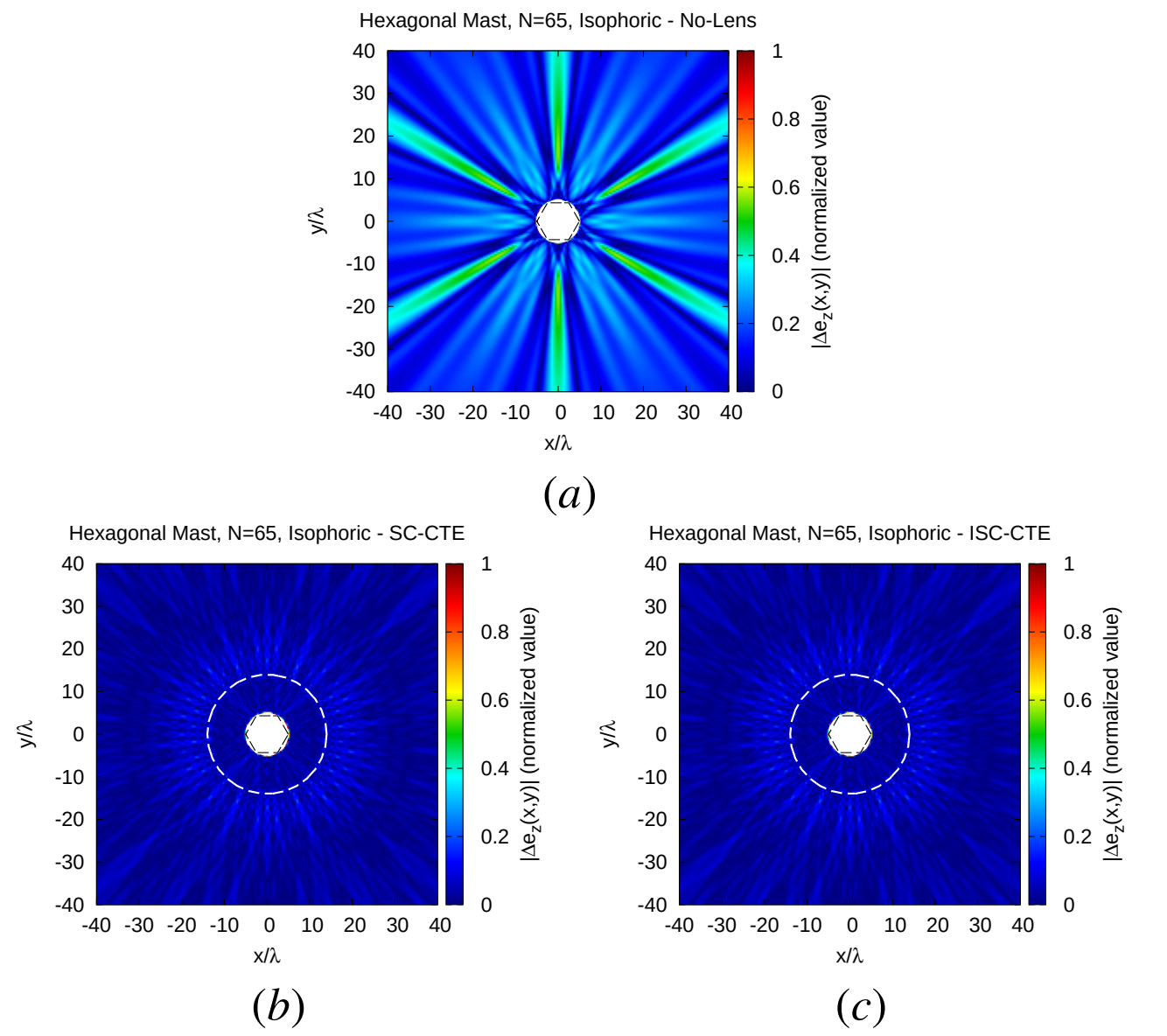

Figure 13. *Numerical Validation* ("*Hexagonal Mast*", $N = 65$, $P = 6$, $M = 30$) - Plot of the magnitude of the normalized difference field, $|\Delta e_z(\mathbf{r})|$, when the reference array radiates in (*a*) in free-space and in the presence of (*b*) the *SC-CTE* and (*c*) the *ISC-CTE* metalenses.

Table II
*Numerical Validation* ($N = 65$) - ANISOTROPY INDEXES FOR THE "*Square Mast*" (FIG. 3) AND THE "*Hexagonal Mast*" (FIG. 12) TEST CASES.

| Test Case | *SC-CTE* | | *QCTE* | |
|---|---|---|---|---|
| | $\alpha_F$ | $\alpha_R$ | $\alpha_F$ | $\alpha_R$ |
| Fig. 3 | $1.94 \times 10^{-1}$ | $1.56 \times 10^{-1}$ | $2.73 \times 10^{-1}$ | $2.39 \times 10^{-1}$ |
| Fig. 12 | $1.03 \times 10^{-1}$ | $8.93 \times 10^{-2}$ | $1.33 \times 10^{-1}$ | $1.18 \times 10^{-1}$ |

For completeness, the *SC-CTE FMD* performance are compared with those of the *QCTE* alternative one when exciting the 16 array elements ($N = 65$) on the *top region* of the contour [$N_1$-arrangement as shown in the inset of Fig. 14(*a*)] or in one of the surface corners [$N_2$-arrangement as in the inset of Fig. 14(*b*)]. As expected, also from the outcomes of the same analysis carried out for the "square-mast", the *QCTE*-based solution under performs with respect to the *SC-CTE* device [e.g., $\eta^{QCTE}\rfloor_{N_1} \approx 2.01\%$ vs. $\eta^{SC-CTE}\rfloor_{N_1} \approx 0.93\%$ - Fig. 14(*a*)] even more in the "corner excitation" case [e.g., $\eta^{QCTE}\rfloor_{N_2} \approx 16.23\%$ vs. $\eta^{SC-CTE}\rfloor_{N_2} \approx 2.05\%$ - Fig. 14(*b*)] since the beam energy is focused along the *permittivity peak*-regions [Fig. 15(*a*) vs. Fig. 15(*b*)]. Dealing with the "isotropic" implementation of the lens, the superiority of the *ISC-CTE FMD* is very evident as indicated by the values of the matching error: $\eta^{IQCTE}\rfloor_{N_1} \approx 70.31\%$ vs. $\eta^{ISC-CTE}\rfloor_{N_1} \approx 1.01\%$ [Fig. 14(*a*)] and $\eta^{IQCTE}\rfloor_{N_2} \approx 80.11\%$ vs. $\eta^{ISC-CTE}\rfloor_{N_2} \approx 3.01\%$ [Fig. 14(*b*)]. This is clear not only at the "integral" level (i.e., $\eta$ values), but also at the "local" level of the field mismatch values. As a matter of fact, it turns out that $\left|\Delta e_z^{IQCTE}(\mathbf{r})\right|_{N_1} \leq 6.95 \times 10^{-1}$ vs. $\left|\Delta e_z^{ISC-CTE}(\mathbf{r})\right|_{N_1} \leq 1.60 \times 10^{-1}$ [Fig. 15(*c*) vs. Fig. 15(*e*)] and $\left|\Delta e_z^{IQCTE}(\mathbf{r})\right|_{N_2} \leq 9.31 \times 10^{-1}$ vs. $\left|\Delta e_z^{ISC-CTE}(\mathbf{r})\right|_{N_2} \leq 2.17 \times 10^{-1}$ [Fig. 15(*d*) vs. Fig. 15(*f*)]. The possibility of the *SC-CTE* method to synthesize an effective and reliable *FMD* also when enforcing supplementary, be-yond the isotropic approximation in (25), lens simplifications has been then assessed. Towards this purpose and following the *tiling* approach discussed in [19], the dielectric profile $\varepsilon^{iso}(\mathbf{r})$ of the isotropic lens in Fig. 12(*f*) has been radially tessellated in $N_\rho \times N_\varphi$ ($N_\rho = \frac{\nu_o - \nu_i}{\Delta_\ell}$ and $N_\varphi = \frac{2\pi\nu_o}{\Delta_\ell}$, $\Delta_\ell$ being the discretization step) homogeneous regions. The analysis of $|\Delta e_z(\mathbf{r})|$ when isophorically exciting all the $N = 65$ array elements indicates that, as expected, the effectiveness of the *FMD* with the tiled lens worsen as coarser discretizations are used (e.g., $|\Delta e_z(\mathbf{r})|_{\Delta_\ell = 0.5\,\lambda} \leq 2.11 \times 10^{-1}$ [Fig. 16(*b*)], $|\Delta e_z(\mathbf{r})|_{\Delta_\ell = 1.0\,\lambda} \leq 2.23 \times 10^{-1}$ [Fig. 16(*d*)], and $|\Delta e_z(\mathbf{r})|_{\Delta_\ell = 2.0\,\lambda} \leq 4.14 \times 10^{-1}$ [Fig. 16(*f*)]), but only when $\Delta_\ell \geq 2.0\,\lambda$ the *SC-CTE FMD* starts loosing its beam-focusing properties by generating multiple-beams as shown in Fig. 16(*f*). This latter behavior arises earlier (i.e., $\Delta_\ell \leq 1.0\,\lambda$) when considering the "top excitation" scenario [Fig. 17(*a*)] as quantitatively confirmed by the $\eta$ values in Fig. 17(*b*) [e.g., $\eta_{\Delta=\lambda} \approx 5\%$). While further investigations, also taking into account various manufacturing processes, are needed to reach reliable operative guidelines, these outcomes are a preliminary proof that the *SC-CTE* technique can be profitably exploited to design double-connected lenses with *isotropic* and *tiled* compositions, thus, in turn, it can potentially enable simpler fabrication procedures [10].

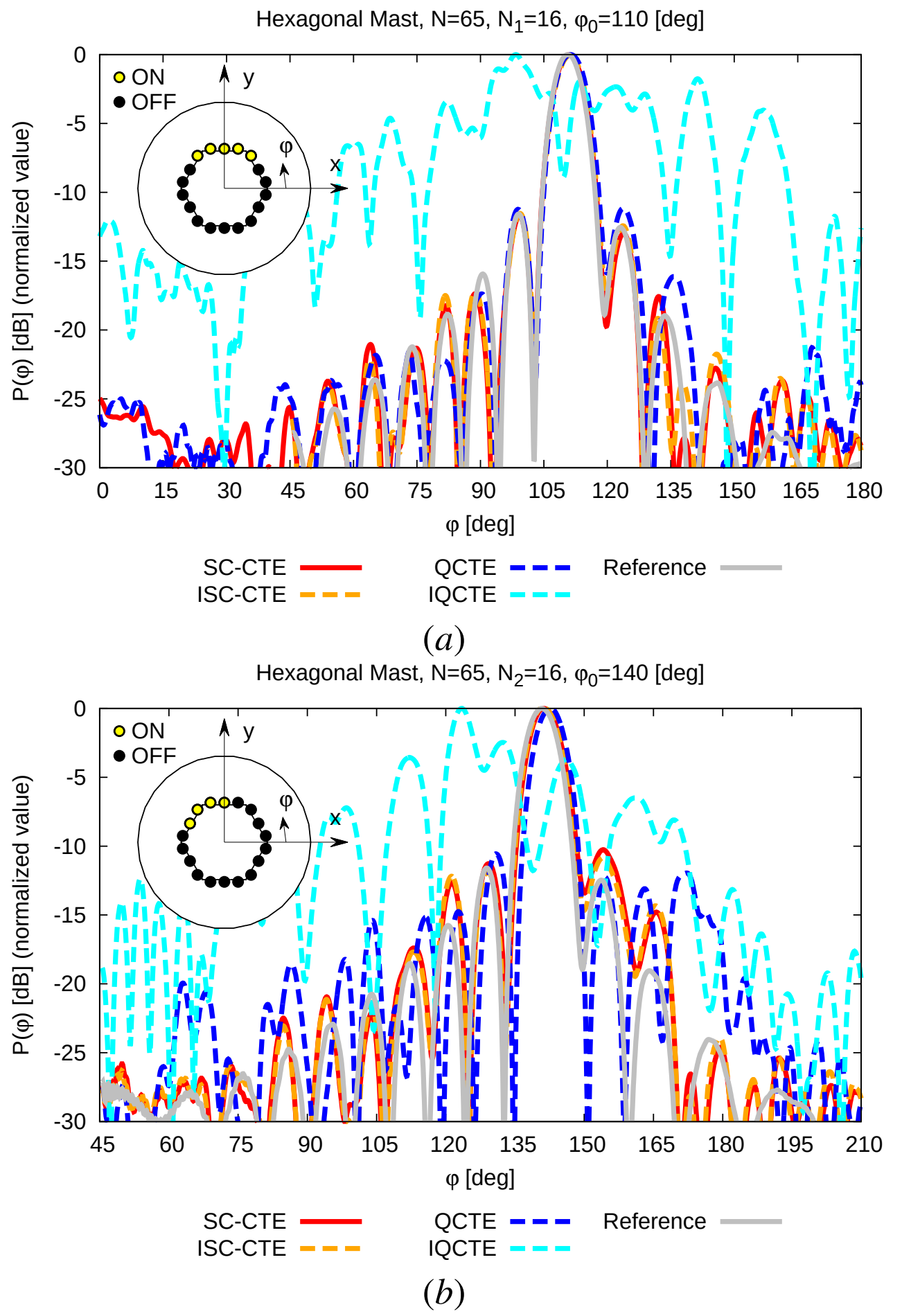

Figure 14. *Numerical Validation* ("*Hexagonal Mast*", $N = 65$, $P = 6$, $M = 30$) - Plot of the normalized power pattern, $\mathcal{P}(\varphi)$, of the reference array and of the *FMD*s with the *SC-CTE* and the *QCTE* lenses when exciting (*a*) the top ($N_1 = 16$) array elements to steer the beam along $\varphi_0 = 110$ [deg] and (*b*) the corner ($N_2 = 16$) array elements to steer the beam towards $\varphi_0 = 140$ [deg].

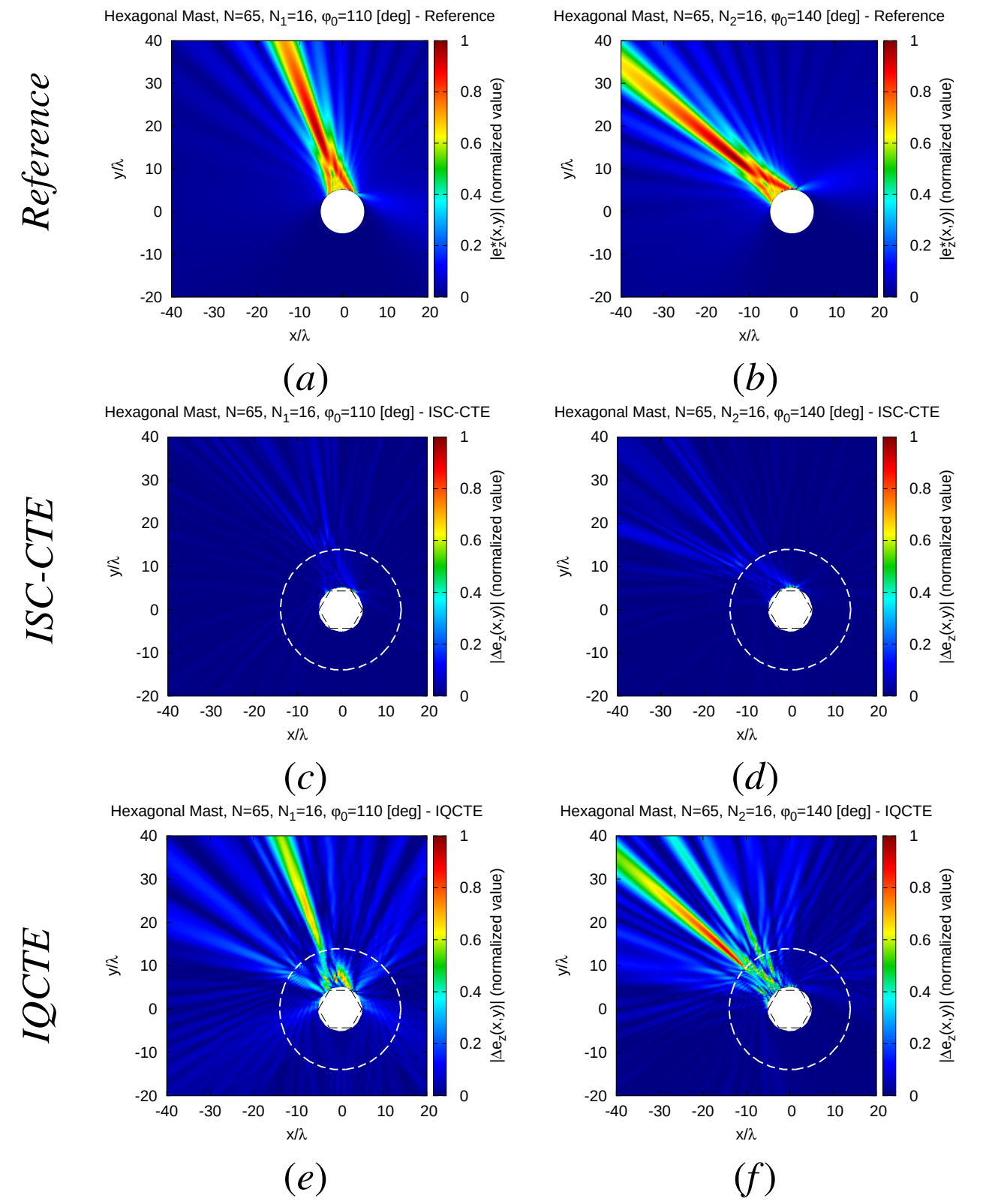

Figure 15. *Numerical Validation* ("*Hexagonal Mast*", $N = 65$, $P = 6$, $M = 30$) - Plot of the magnitude of (*a*)(*b*) the target field distribution, $e^*(\mathbf{r})$, and of (*c*)-(*f*) the normalized difference field,$|\Delta e_z(\mathbf{r})|$, for the (*b*)(*d*) *ISC-CTE* and (*c*)(*e*) *IQCTE FMD*s when exciting (*a*) the top ($N_1 = 16$) array elements to steer the beam along $\varphi_0 = 110$ [deg] and (*b*) the corner ($N_2 = 16$) array elements to steer the beam towards $\varphi_0 = 140$ [deg].

Table III
*Numerical Validation* ("*Irregular Masts*", $N = 65$) - ANISOTROPY INDEXES.

| Test Case | $\alpha_F$ | $\alpha_R$ |
|---|---|---|
| Fig. 18(*a*) | $1.70 \times 10^{-1}$ | $1.36 \times 10^{-1}$ |
| Fig. 18(*b*) | $2.03 \times 10^{-1}$ | $1.86 \times 10^{-1}$ |

As for the computational efficiency of the *SC-CTE* paradigm, it is worth pointing out that the overall time[4] required to synthesize the *FMD* turns out very limited for any $\Delta_\ell$ value [i.e., $\Delta t \leq 10.5$ [s] - Fig. 17(*b*)]. Such a result is actually expected from the theoretical viewpoint because of the adoption of a local search technique for the solution of (29) (see the Appendix), which guarantees very effective convergence performance regardless of the required transformation accuracy.

The final experiment is aimed at assessing the flexibility of the *SC-CTE* synthesis approach in designing *FMD*s conformal to irregular (non-canonical) mast geometries. For instance, the internal contours, $\partial_i\Gamma$, in Fig. 18(*a*) ("*irregular heptagon*") and Fig. 18(*b*) ("*irregular octagon*") have been used as bench-mark examples by keeping the remaining setup of the other examples [i.e., $N = 65$ and the target field in Fig. 2$(a)$]. Once again, the synthesized *SC-CTE* lenses present isotropic dielectric distributions [Figs. 18$(a)$-18$(b)$ and Tab. III] with permittivity peaks in correspondence with the sharpest edges of the inner surface [e.g., Fig. 18$(b)$]. Moreover, both the near ["*irregular heptagon*" - Fig. 18$(c)$; "*irregular octagon*" - Fig. 18$(d)$] and the far-field, $\mathcal{P}(\varphi)$ ["*irregular heptagon*" - Fig. 19$(a)$; "*irregular octagon*" - Fig. 19$(b)$], field distributions, radiated when isophorically exciting the whole array, assess the effectiveness of the *SC-CTE* technique since $|\Delta e_z(\mathbf{r})| \leq 2.82 \times 10^{-1}$ [Fig. 18$(c)$] and $|\Delta e_z(\mathbf{r})| \leq 2.26 \times 10^{-1}$ [Fig. 18$(d)$] as well as $\frac{\eta^{ISC-CTE}}{\eta^{No-Lens}} \approx 4.11 \times 10^{-2}$ [Fig. 19$(a)$] and $\frac{\eta^{ISC-CTE}}{\eta^{No-Lens}} \approx 8.72 \times 10^{-2}$ [Fig. 19$(b)$].

[4]For the sake of fairness, all simulations have been carried out exploiting non-optimized MATLAB codes executed on a single-core CPU running at 2.6 GHz.

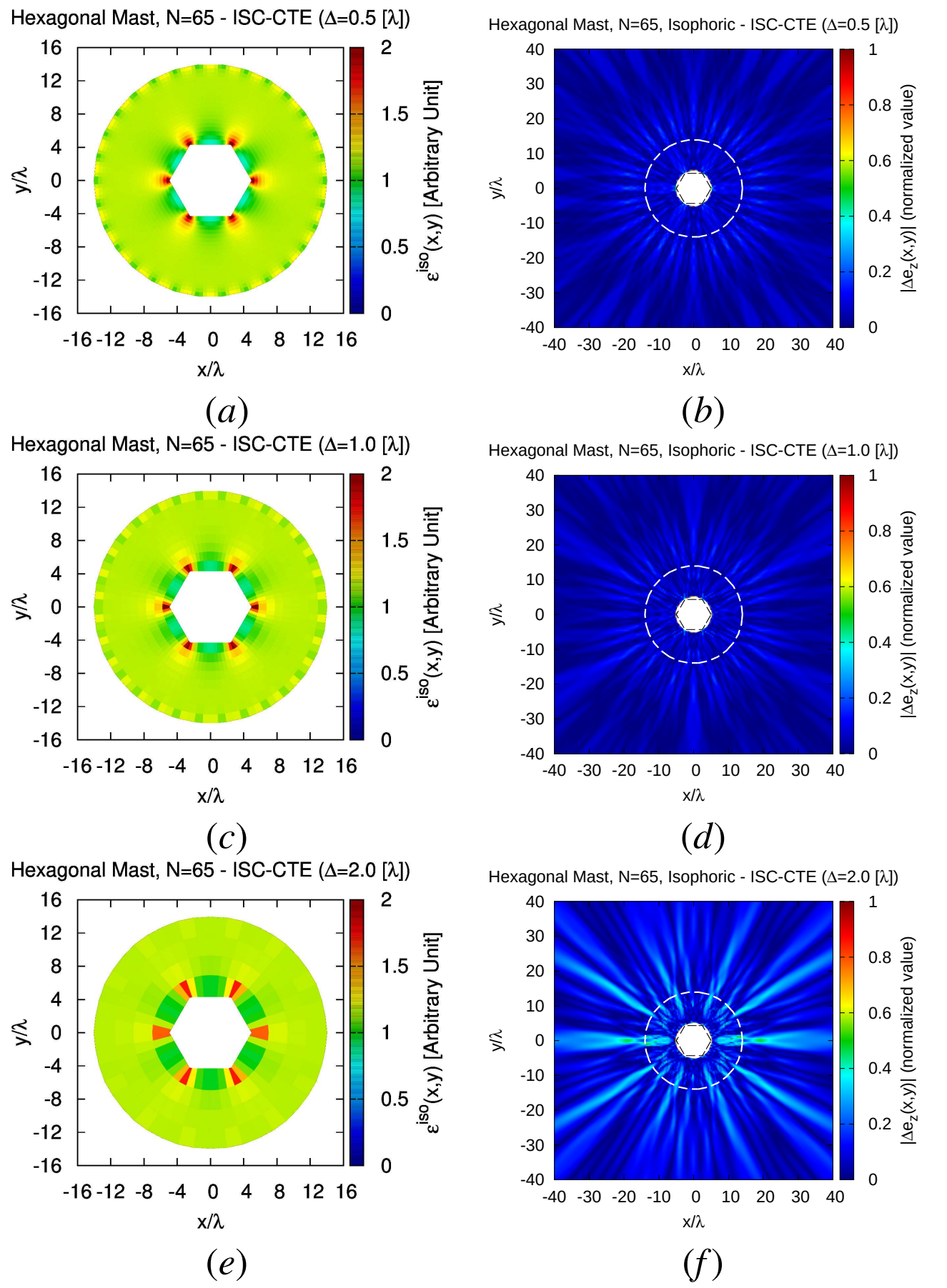

Figure 16. *Numerical Validation* ("*Hexagonal Mast*", $N = 65$, $P = 6$, $M = 30$) - Plots of $(a)(c)(e)$ the permittivity distribution and of $(b)(d)(f)$ the magnitude of the normalized difference field,$|\Delta e_z(\mathbf{r})|$, for a tiled *ISC-CTE FMD* with discretization step equal to $(a)(b)$ $\Delta_\ell = 0.5$ $[\lambda]$, $(c)(d)$ $\Delta_\ell = 1.0$ $[\lambda]$, and $(e)(f)$ $\Delta_\ell = 2.0$ $[\lambda]$.

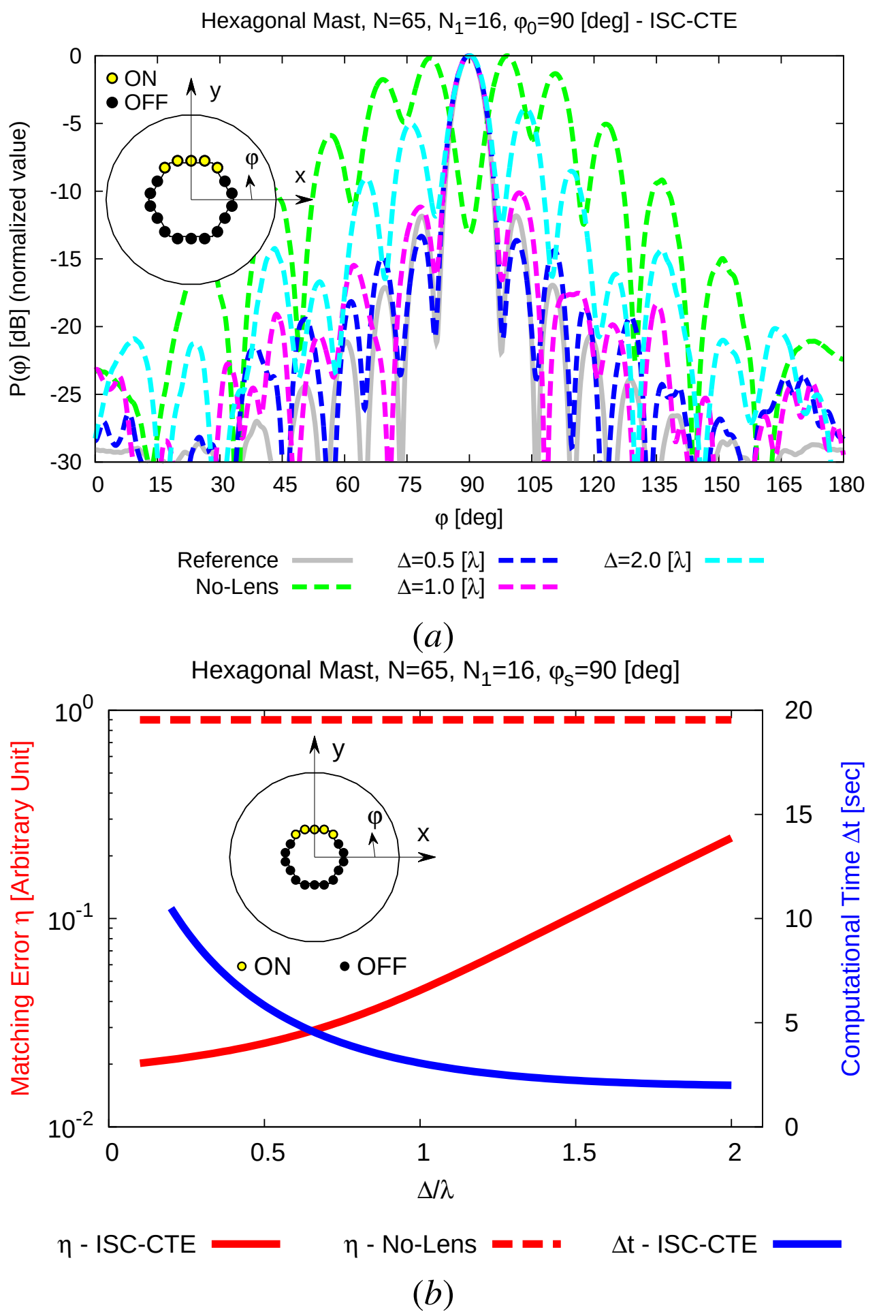

Figure 17. *Numerical Validation* ("*Hexagonal Mast*", $N = 65$, $P = 6$, $M = 30$ - $N_1 = 16$, $\varphi_0 = 90$ [deg]) - Plot of $(a)$ the normalized power pattern, $\mathcal{P}(\varphi)$, of the reference array and of the *ISC-CTE FMD*s tiled with different discretization steps ($\Delta_\ell \in [0.1, 2.0]$ $[\lambda]$), and of $(b)$ $\eta$ and $\Delta t$ versus $\Delta_\ell$.

## V. Conclusions and Remarks

A new *TE*-based synthesis strategy, which leverages on the Schwarz-Christoffel theorem, has been proposed. Owing to the features of the *SC* mapping formula (16), such a method theoretically avoids/minimizes the anisotropy of the arising metalenses thanks to the conformal nature of the transformation function. Moreover, arbitrary doubly-connected domains can be effectively handled, thus enabling the development of *FMD*s installed on masts with arbitrary cross-sections. Selected experiments drawn from a wide numerical validation have been presented to point out the advantages and the potentialities of the *SC-CTE* synthesis approach in comparison with state-of-the-art *TE* approaches, as well.

The numerical assessment has shown that

- the proposed approach yields lenses with uniaxially-anisotropic materials (e.g., Fig. 3), whose isotropic simplifications similarly performs [e.g., Fig. 5$(a)$];
- both the *SC-CTE* and the *ISC-CTE* implementations carefully match the target field distributions regardless of the geometry and the excitations/steering at hand (e.g., Fig. 7 and Fig. 18);
- *SC*-based *FMD*s positively compare with the layouts derived with state-of-the-art *TE* methods in terms of lens anisotropy (Tab. II) and field control capabilities (Fig. 9 - Fig. 14) showing that the choice of *SC* mapping potentially yields more effective metalens designs while also enabling holes/empty regions within the synthesis procedure;
- besides the isotropic approximation, further lens sim-

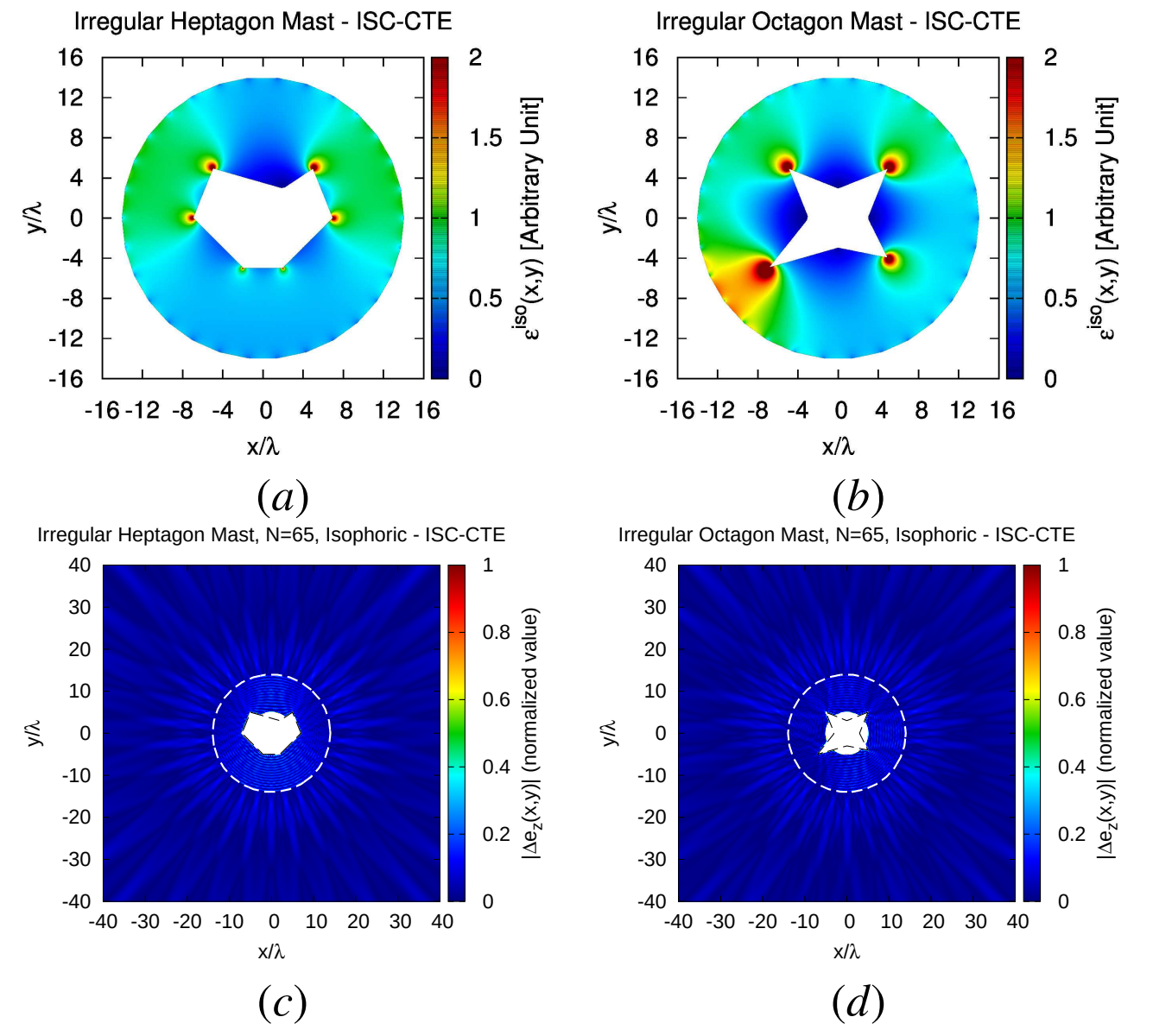

Figure 18. *Numerical Validation* ("*Irregular Masts*", $N = 65$) - Plots of $(a)(b)$ the permittivity distribution and of $(b)(d)(f)$ the magnitude of the normalized difference field, $|\Delta e_z(\mathbf{r})|$, of the *ISC-CTE FMD*s synthesized for the $(a)(c)$ the *irregular heptagon* and $(b)(d)$ the *irregular octagon* mast geometries.

plifications of the *SC-CTE* lenses, such as those based on material *tiling*, turn out to be satisfactory (Fig. 17) because of the bounded and low-amplitude peaks in the synthesized permittivity distributions.

As for the main methodological advancements with respect to the state-of-the-art, they include $(i)$ the customization of the *SC* mapping and its application to the solution of *TE* synthesis problems concerned with the manipulation of the *EM* radiation of arrays, $(ii)$ the derivation of a *TE* paradigm suitable for handling doubly-connected lenses such as those for masts and/or for supports/domains with forbidden regions, and $(iii)$ the definition of operative guidelines for a reliable application of the proposed *SC*-based method for the synthesis of *FMD*s.

Future works, out-of-the-scope of this paper, will include the generalization of the *SC-CTE* formulation to other geometrical setups. The experimental implementation/validation of *SC-CTE* designed *FMD*s is currently under investigation since beyond the possibilities of our current fabrication/measurement facilities.

## Appendix

### *Determination of the* SC *Accessory Parameters in (16)*

According to *SC* theory, the mapping function $\psi\left(\widetilde{\zeta}\right)$ is uniquely determined up to one degree of freedom, which is usually eliminated by assuming $\widetilde{\zeta}_o^{(M)} = 1.0$ [24][25][30]. The computation of the remaining accessory parameters [i.e., the outer *prevertexes* $\widetilde{\zeta}_o^{(m)}$, $m = 1, ..., M-1$, and the inner *prevertexes* $\widetilde{\zeta}_i^{(p)}$, $p = 1, ..., P$, the complex constant $C$, and the internal radius $\mu$] is carried out by solving the following

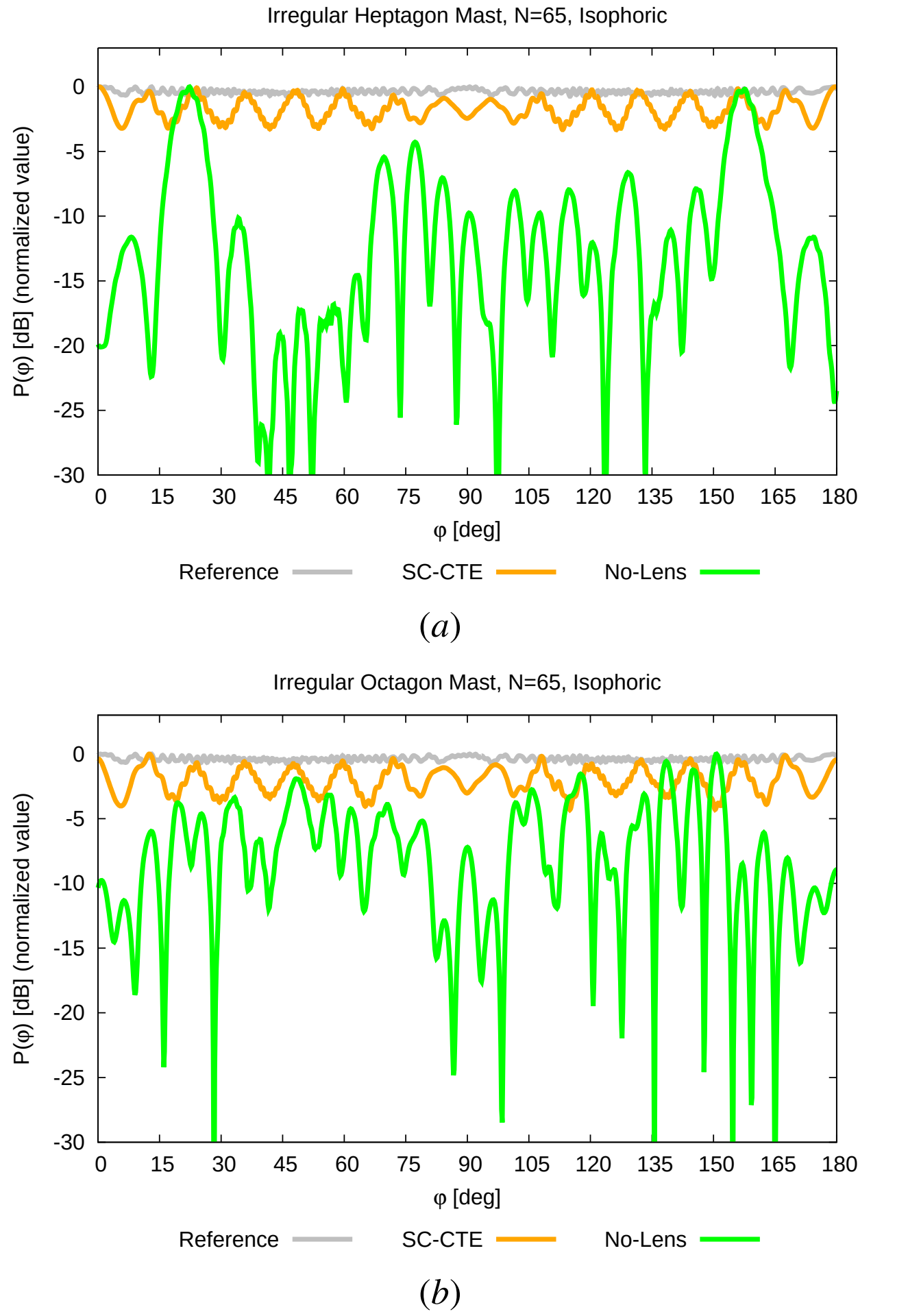

Figure 19. *Numerical Validation* ("*Irregular Masts*", $N = 65$) - Plot of the normalized power pattern, $\mathcal{P}(\varphi)$, of the reference array and of the *ISC-CTE FMD* when dealing with $(a)$ the *irregular heptagon* and $(b)$ the *irregular octagon* mast geometries.

nonlinear system of equations [25]

$$\begin{cases} C\int_{\widetilde{\zeta}_o^{(M)}}^{\widetilde{\zeta}_o^{(1)}} \mathcal{Q}(s)\,\mathrm{d}s = \zeta_o^{(1)} - \zeta_o^{(M)} & \\ \left|C\int_{\widetilde{\zeta}_o^{(m)}}^{\widetilde{\zeta}_o^{(m+1)}} \mathcal{Q}(s)\,\mathrm{d}s\right| = \left|\zeta_o^{(m+1)} - \zeta_o^{(m)}\right| & m = 1, ..., M-3 \\ C\int_{\widetilde{\zeta}_i^{(P)}}^{\widetilde{\zeta}_i^{(1)}} \mathcal{Q}(s)\,\mathrm{d}s = \zeta_i^{(1)} - \zeta_i^{(P)} & \\ \left|C\int_{\widetilde{\zeta}_i^{(p)}}^{\widetilde{\zeta}_i^{(p+1)}} \mathcal{Q}(s)\,\mathrm{d}s\right| = \left|\zeta_i^{(p+1)} - \zeta_i^{(p)}\right| & p = 1, ..., P-1 \\ C\int_{\widetilde{\zeta}_o^{(M)}}^{\widetilde{\zeta}_i^{(P)}} \mathcal{Q}(s)\,\mathrm{d}s = \zeta_i^{(P)} - \zeta_o^{(M)} & \end{cases} \tag{28}$$

taking into account the constraints $(i)$ $\mu \in (0, 1)$, $(ii)$ $\angle\widetilde{\zeta}_o^{(m)} < \angle\widetilde{\zeta}_o^{(m+1)}$, $m = 1, ..., M-1$, and $(iii)$ $\angle\widetilde{\zeta}_i^{(p)} < \angle\widetilde{\zeta}_i^{(p+1)}$, $p = 1, ..., P-1$, where $\angle\cdot$ stands for the complex number argument. The solution of the *SC* system (28) is numerically not trivial because of the presence of singularities in $\mathcal{Q}(s)$ and the (possible) occurrence of numerical instabilities due to the integration path [25]. To address these challenges, the definite integrals in (28) are $(i)$ rewritten according to the Daeppen change of variables [25], $(ii)$ discretized with a Gauss-Jacobi quadrature technique, and $(iii)$ solved by choosing - whenever possible - circular-arc integration paths to achieve the numerical stability of the integration [25]. The discretized system of equations is finally solved, in the least square sense, as the minimization of the cost function $\Phi$

$$\Phi\left(C;\mu;\widetilde{\zeta}_i^{(p)},\,p=1,....P;\,\widetilde{\zeta}_o^{(m)},\,m=1,...,M-1\right)\triangleq \left|C\int_{\widetilde{\zeta}_o^{(M)}}^{\widetilde{\zeta}_o^{(1)}}\mathcal{Q}\left(s\right)\mathrm{d}s-\zeta_o^{(1)}+\zeta_o^{(M)}\right|^2+ \sum_{m=1}^{M-3}\left|\left|C\int_{\widetilde{\zeta}_o^{(m)}}^{\widetilde{\zeta}_o^{(m+1)}}\mathcal{Q}\left(s\right)\mathrm{d}s\right|-\left|\zeta_o^{(m+1)}-\zeta_o^{(m)}\right|\right|^2+ \left|C\int_{\widetilde{\zeta}_i^{(P)}}^{\widetilde{\zeta}_i^{(1)}}\mathcal{Q}\left(s\right)\mathrm{d}s-\zeta_i^{(1)}+\zeta_i^{(P)}\right|^2+ \sum_{p=1}^{P-1}\left|\left|C\int_{\widetilde{\zeta}_i^{(p)}}^{\widetilde{\zeta}_i^{(p+1)}}\mathcal{Q}\left(s\right)\mathrm{d}s\right|-\left|\zeta_i^{(p+1)}-\zeta_i^{(p)}\right|\right|^2+ \left|C\int_{\widetilde{\zeta}_o^{(M)}}^{\widetilde{\zeta}_i^{(P)}}\mathcal{Q}\left(s\right)\mathrm{d}s-\zeta_i^{(P)}+\zeta_o^{(M)}\right|^2. \tag{29}$$

with respect to the *SC* accessory parameters.
Towards this end and following the guidelines in [25], an iterative gradient-based minimization technique based on the Hybrid Powell method [31] is adopted.

## *Authors’ Information*

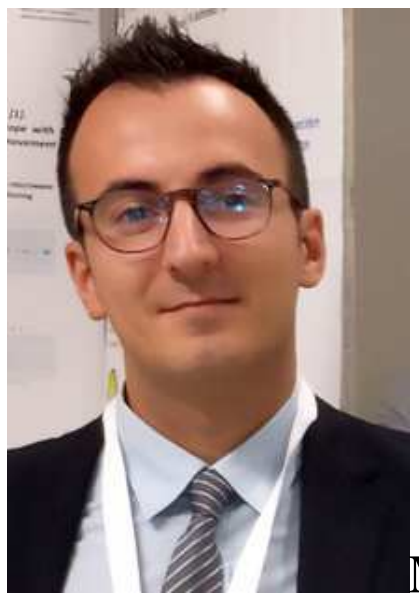

**Marco Salucci** (GS’13-M’15) received the M.S. degree in Telecommunication Engineering from the University of Trento, Italy, in 2011, and the Ph.D. degree from the International Doctoral School in Information and Communication Technology of Trento in 2014. From December 2014 to November 2016 he was a Post Doc Researcher at CentraleSupélec, in Gif-sur-Yvette, France. From December 2016 to September 2017 he was a Post Doc Researcher at the Commissariat à l'énergie atomique et aux énergies alternatives (CEA), in Gif-sur-Yvette, France. He is currently a Researcher at the Department of Information Engineering and Computer Science (DISI) at the University of Trento, Italy, and a member of the ELEDIA Research Center. Dr. Salucci is a Member and of the IEEE Antennas and Propagation Society and he was a Member of the COST Action TU1208 "Civil Engineering Applications of Ground Penetrating Radar". He serves as an Associate Editor of the IEEE Transactions on Antennas and Propagation and as a reviewer for different international journals including IEEE Antennas and Wireless Propagation Letters, IEEE Journal on Multiscale and Multiphysics Computational Techniques, and IET Microwaves, Antennas & Propagation. His research activities are mainly concerned with inverse scattering, GPR microwave imaging techniques, antenna synthesis, and computational electromagnetics with focus on System-by-Design methodologies integrating optimization techniques and Learning-by-Examples methods for real-world applications.

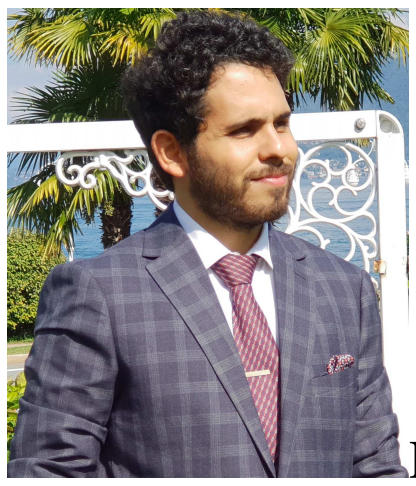

**Federico Boulos** received the MSc in Information and Communications Engineering in 2019 and BSc in Telecommunications and Electronics Engineering in 2017 both at the University of Trento, Italy. He is a member of the ELEDIA Research Center and he is currently a Ph.D. student of the ICT International Doctoral School of Trento. His research activity is mainly focused on the synthesis of non-conventional antennas and metamaterial-enhanced devices based on the SbD paradigm.

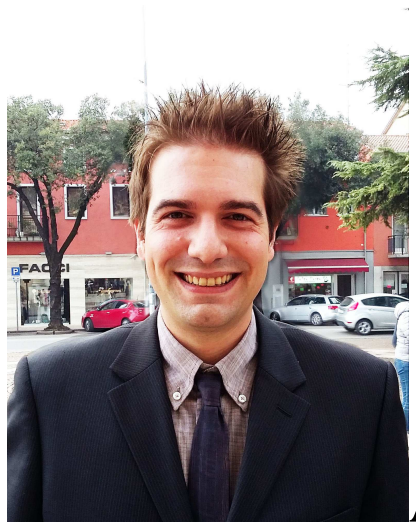

**Alessandro Polo** (GS'13-M'14) received the M.S. degree in Telecommunication Engineering from the University of Trento (Italy) in 2012 and the PhD degree from the International Doctoral School in Information and Communication Technology of Trento in 2018. He is currently an PostDoc Researcher at the Department of Information Engineering and Computer Science (University of Trento) and a senior member of the ELEDIA Research Center. His research work is focused mainly on wireless networking, distributed computing, localization, decision support systems for smart city, fleet and critical applications management. He has a long experience in the fields of KPI analysis and troubleshooting, machine learning, multi-objective evolutionary techniques and software development.

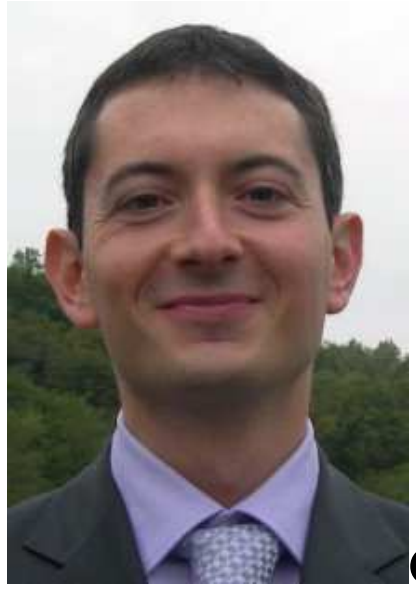

**Giacomo Oliveri** (GS'06-M'09-SM'13) received the B.S. and M.S. degrees in Telecommunications Engineering and the PhD degree in Space Sciences and Engineering from the University of Genoa, Italy, in 2003, 2005, and 2009 respectively. He is currently an Associate Professor at the Department of Information Engineering and Computer Science (University of Trento) and a member of the ELEDIA Research Center. Moreover, he is Adjunct Professor at CentraleSupélec and member of the Laboratoire des signaux et systèmes (L2S)@CentraleSupélec Gif-sur-Yvette (France). He has been a visiting researcher at L2S in 2012, 2013, and 2015, Invited Associate Professor at the University of Paris Sud, France, in 2014, and visiting professor at Université Paris-Saclay in 2016 and 2017.
He is author/co-author of over 330 peer reviewed papers on international journals and conferences. His research work is mainly focused on electromagnetic direct and inverse problems, system-by-design and metamaterials, and antenna array synthesis. Prof. Oliveri serves as an Associate Editor of the IEEE Antennas and Wireless Propagation Letters, of the IEEE Journal on Multiscale and Multiphysics Computational Techniques, of the International Journal of Antennas and Propagation, of the International Journal of Distributed Sensor Networks, and of the Microwave Processing journal. He is a Senior Member of the IEEE, and the Chair of the IEEE AP/ED/MTT North Italy Chapter.